\pdfoutput=1
\documentclass[iop]{emulateapj}

\slugcomment{}

\shorttitle{Evidence of interacting elongated filaments in the star-forming site AFGL 5142}
\shortauthors{L.~K. Dewangan et al.}
\begin{document}

\title{Evidence of interacting elongated filaments in the star-forming site AFGL 5142}
\author{L.~K. Dewangan\altaffilmark{1}, D.~K. Ojha\altaffilmark{2}, T. Baug\altaffilmark{3}, and R. Devaraj\altaffilmark{4}}
\email{Email: lokeshd@prl.res.in}
\altaffiltext{1}{Physical Research Laboratory, Navrangpura, Ahmedabad - 380 009, India.}
\altaffiltext{2}{Department of Astronomy and Astrophysics, Tata Institute of Fundamental Research, Homi Bhabha Road, Mumbai 400 005, India.}
\altaffiltext{3}{Kavli Institute for Astronomy and Astrophysics, Peking 
University, 5 Yiheyuan Road, Haidian District, Beijing 100871, P. R. China.}
\altaffiltext{4}{Instituto Nacional de Astrof\'{\i}sica, \'{O}ptica y Electr\'{o}nica, Luis Enrique Erro \# 1, Tonantzintla, Puebla, M\'{e}xico C.P. 72840.}
\begin{abstract}
To probe the ongoing physical mechanism, we studied a wide-scale environment around AFGL 5142 (area $\sim$25 pc $\times$ 20 pc) using a multi-wavelength approach. 
The {\it Herschel} column density ($N(\mathrm H_2)$) map reveals a massive inverted Y-like structure (mass $\sim$6280 M$_{\odot}$), which hosts a pair of elongated filaments (lengths $>$ 10 pc). 
The {\it Herschel} temperature map depicts the filaments in a temperature range of $\sim$12.5--13.5~K. 
These elongated filaments overlap each other at several places, where $N(\mathrm H_2)$ $>$ 4.5 $\times$ 10$^{21}$ cm$^{-2}$. 
The $^{12}$CO and $^{13}$CO line data also show two elongated cloud components (around $-$1.5 and $-$4.5 km s$^{-1}$) toward 
the inverted Y-like structure, which are connected in the velocity space. 
First moment maps of CO confirm the presence of two intertwined filamentary clouds along the line of sight. 
These results explain the morphology of the inverted Y-like structure through a combination of two different filamentary clouds, which are also supported by the distribution of the cold H\,{\sc i} gas. Based on the distribution of young stellar objects (YSOs), star formation (SF) activities are investigated toward the 
inverted Y-like structure.
The northern end of the structure hosts AFGL 5142 and tracers of massive SF, where high surface density 
of YSOs (i.e., 5--240 YSOs pc$^{-2}$) reveals strong SF activity. 
Furthermore, noticeable YSOs are found toward the overlapping zones of the clouds. 
All these observational evidences support a scenario of collision/interaction of two elongated filamentary clouds/flows, which appears to explain SF history in the site AFGL 5142.
\end{abstract}
\section{Introduction}
\label{sec:intro}
Numerous observational investigations, using the infrared and sub-millimeter (sub-mm) data, have revealed the existence of filamentary structures at different length scales in star-forming sites \citep[e.g.,][]{andre14}, which are very 
important ingredients in the formation process of young stellar clusters and massive stars. 
Currently, the investigation of the role of filaments in the birth of dense 
massive star-forming clumps and young stellar clusters is one of the ongoing research topics in the area of star formation (SF) \citep[e.g.,][]{myers09,andre10,andre14,schneider12,nakamura14,baug15,dewangan15,dewangan16,dewangan16b,dewangan17a,
dewangan17b,dewangan17c,dewangan17d,dewangan17e,contreras16,liu16,williams18}. 

AFGL 5142 is an active site of SF inferred by the presence of multiple outflows on multiple scales \citep{hunter99,zhang07,palau11,palau13}, and the detection of various masers \citep{hunter95,goddi06,qiu08,goddi11,liu16}. 
The site is located at a distance of 2.14 kpc \citep{burns17}, and hosts a group of millimeter (mm) cores \citep{zhang07,palau11,palau13}. 
One of the mm cores \citep[i.e., MM1; M$_{core}$ $\sim$6.5 M$_{\odot}$;][]{liu16} 
contains the 6.7-GHz methanol masers, favouring the presence of at least one embedded massive star in the AFGL 5142 MM1 system \citep{goddi11}.
\citet{goddi11} reported that the core AFGL 5142 MM1 shows radio continuum emission from ionized gas, 
and excites the 22-GHz water maser and the 6.7-GHz methanol maser. Furthermore, another mm core, MM2 \citep[M$_{core}$ $\sim$6.2 M$_{\odot}$;][]{liu16} is investigated, where no radio continuum/centimeter emission or 6.7-GHz methanol maser is detected. 
However, the presence of the 22-GHz water maser in the direction of the AFGL 5142 MM2 
system \citep{hunter95,goddi06} indicates the onset of SF.
Using the {\it Spitzer} infrared observations, \citet{qiu08} examined the infrared counterpart 
of the detected mm core, and found an embedded object, which was detected only in the image at 24 $\mu$m (see Table~5 in their paper). 
With the knowledge of chemical compositions, previous observations revealed the signatures of hot molecular core toward the MM1 and MM2 \citep{zhang07,palau11}. \citet{hunter99} presented the HCO$^{+}$(1--0) spectrum of AFGL 5142, which showed at least two velocity peaks around $-$4.5 and $-$1.5 km s$^{-1}$ \citep[see also][]{liu16}. 
In a velocity range of [$-$10, 0] km s$^{-1}$, \citet{colzi18} also observed two peaks toward AFGL 5142-MM using 
the H$^{13}$CN profile. Using the HCN (3--2) and HCO$^{+}$ (3--2) line data sets, \citet{liu16} also reported an extremely wide-angle bipolar outflow (EWBO) with an opening angle of $\sim$180$^{\circ}$, which was found to be powered from the MM1--MM2 region. Using the NH$_{3}$ (1, 1) line data, they also found a hierarchical network of filaments, and reported the signatures of 
gas inflow along the filaments. These authors suggested the importance of both filamentary inflow and outflow feedback in cluster formation in AFGL 5142. 
However, no attempt is made to examine the filaments and different velocity components on a large-scale, which is important to help 
in understanding the exact ongoing physical process in the site AFGL 5142.
Hence, in this paper, we carry out a detailed study of a large-scale environment (more than 20 pc) around AFGL 5142 using a multi-wavelength approach. 
To understand the physical process in the site AFGL 5142, the present work is focused to examine the distribution of dust (i.e., warm and cold), gas (i.e., neutral, molecular, and ionized), and young stellar objects (YSOs). 
On large-scales, the sub-mm images from the space-based {\it Herschel} telescope facility are employed to investigate the embedded filaments in the site AFGL 5142. The study of internal kinematics of the emitting $^{12}$CO and $^{13}$CO gas in a wide-scale 
environment of AFGL 5142 is also performed. 

In Section~\ref{sec:obser}, the details of the adopted observational datasets are provided. 
In Section~\ref{sec:data}, the identification of filaments and the distribution of dust and gas in the selected site are presented. 
Furthermore, the selection procedures of embedded YSOs in the site AFGL 5142 are also explained in the section. 
Section~\ref{sec:disc} is devoted to explain the possible SF scenario. 
Finally, Section~\ref{sec:conc} provides the main conclusions of this paper.
\section{Data and analysis}
\label{sec:obser}
In this work, the analysis of the observational data sets is carried out for an area of $\sim$0$\degr$.676 $\times$ 0$\degr$.534 
(central coordinates: {\it l} = 174$\degr$.455; {\it b} = $-$0$\degr$.197) hosting the site AFGL 5142.
The data sets in the infrared, sub-mm, and radio regimes were collected from different existing surveys 
(e.g., the Two Micron All Sky Survey \citep[2MASS; $\lambda$ = 1.25, 1.65, 2.2 $\mu$m; resolution $\sim$2$''$;][]{skrutskie06}, 
the UKIRT NIR Galactic Plane Survey \citep[GPS; $\lambda$ = 1.25, 1.65, 2.2 $\mu$m; resolution $\sim$0$''$.8;][]{lawrence07}, 
the Warm-{\it Spitzer} Glimpse360\footnote[1]{http://www.astro.wisc.edu/sirtf/glimpse360/} survey \citep[$\lambda$ = 3.6, 4.5 $\mu$m; resolution $\sim$2$''$;][]{whitney11,benjamin03},
the Wide Field Infrared Survey Explorer\footnote[2]{WISE is a joint project of the
University of California and the JPL, Caltech, funded by the NASA} \citep[WISE; $\lambda$ = 12, 22 $\mu$m; resolution $\sim$6$''$, 12$''$;][]{wright10}, 
the {\it Herschel} Infrared Galactic Plane Survey \citep[Hi-GAL; $\lambda$ = 70, 160, 250, 350, 500 $\mu$m; resolution $\sim$5$''$.8, 12$''$, 18$''$, 25$''$, 37$''$;][]{molinari10,molinari10b}, 
the NRAO VLA Sky Survey \citep[NVSS; $\lambda$ = 21 cm; resolution $\sim$45$\arcsec$;][]{condon98}, and the Canadian Galactic Plane Survey \citep[CGPS; $\lambda$ = 21 cm; resolution $\sim$1$'$ $\times$ 1$'$ csc$\delta$;][]{taylor03}). 
The present work also uses the Five College Radio Astronomy 
Observatory (FCRAO) $^{12}$CO(1-0) and $^{13}$CO(1-0) line data (velocity resolution $\sim$0.25~km\,s$^{-1}$).
The FCRAO beam sizes are $\sim$45$''$ (with angular sampling of 22$\farcs$5) and $\sim$46$''$ (with angular sampling of 22$\farcs$5) 
for $^{12}$CO and $^{13}$CO, respectively. 
Typical rms values for the spectra are 0.25 K for $^{12}$CO and 0.2 K for $^{13}$CO \citep[e.g.,][]{heyer96}.
The selected target field was observed as part of the Extended Outer Galaxy Survey \citep[E-OGS,][]{brunt04}, 
that extends the coverage of the FCRAO Outer Galaxy Survey \citep[OGS,][]{heyer98}. 
\section{Results}
\label{sec:data}
\subsection{Wide-scale physical environment of AFGL 5142}
\label{subsec:u1}
\begin{figure*}
\epsscale{0.85}
\plotone{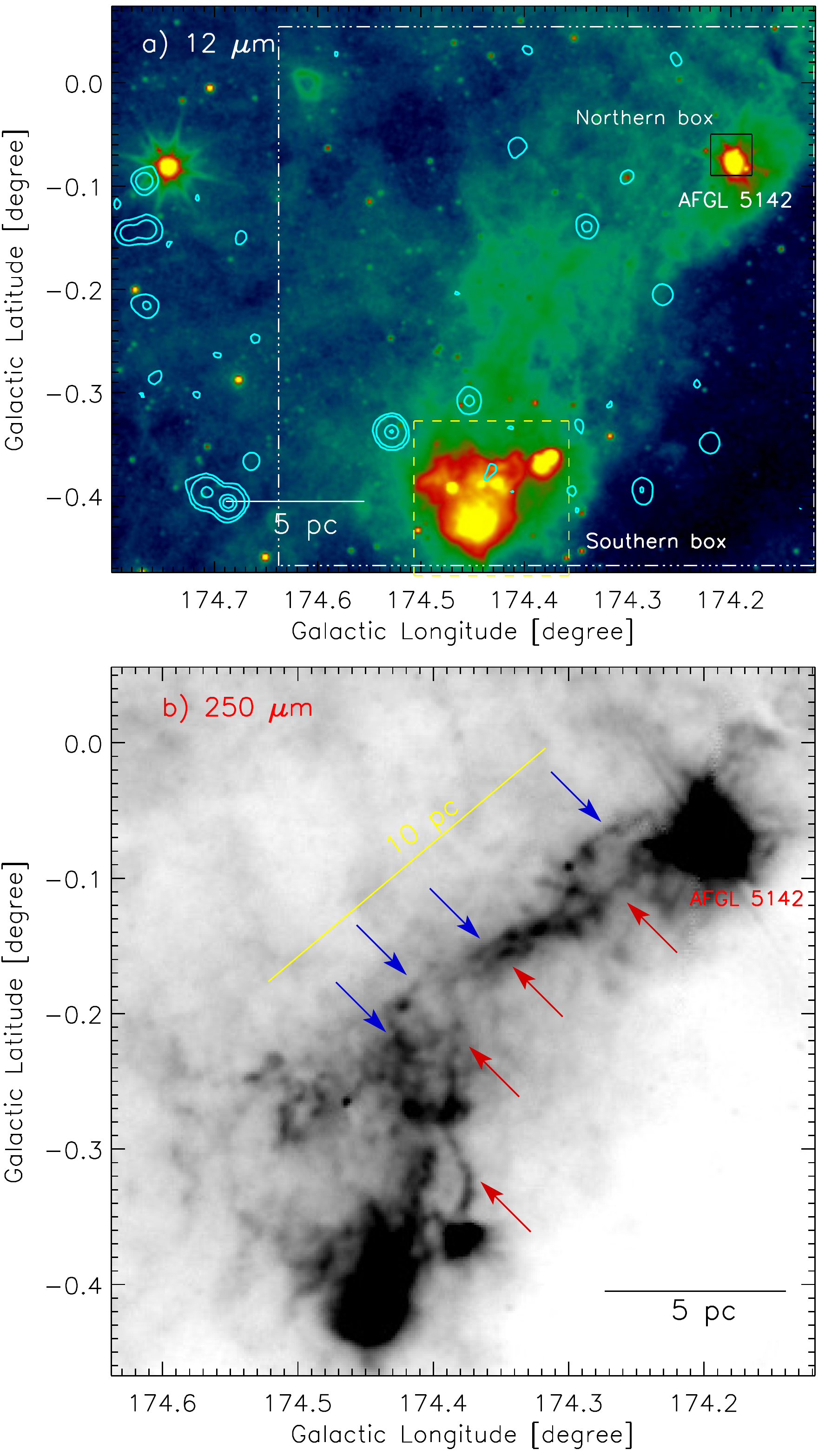}
\caption{a) False color WISE image at 12 $\mu$m of the $\sim$0$\degr$.676 $\times$ 0$\degr$.534 area (central coordinates: {\it l} = 174$\degr$.455; {\it b} = $-$0$\degr$.197) around 
the site AFGL 5142. The image is overlaid with the NVSS radio continuum contours (in cyan) at 1.4 GHz. 
The NVSS contours are shown with the levels of (3, 14, 155, 267) $\times$ 0.45 mJy/beam, where 1$\sigma$ $\sim$0.45 mJy/beam. 
b) Inverted gray-scale {\it Herschel} image at 250 $\mu$m of an area highlighted by a dotted-dashed box (in white) in Figure~\ref{sg1}a. Arrows highlight at least two elongated filaments. 
The scale bar corresponding to 5 pc (at a distance of 2.14 kpc) is displayed in both the panels.}
\label{sg1}
\end{figure*}

\begin{figure*}
\epsscale{1.0}
\plotone{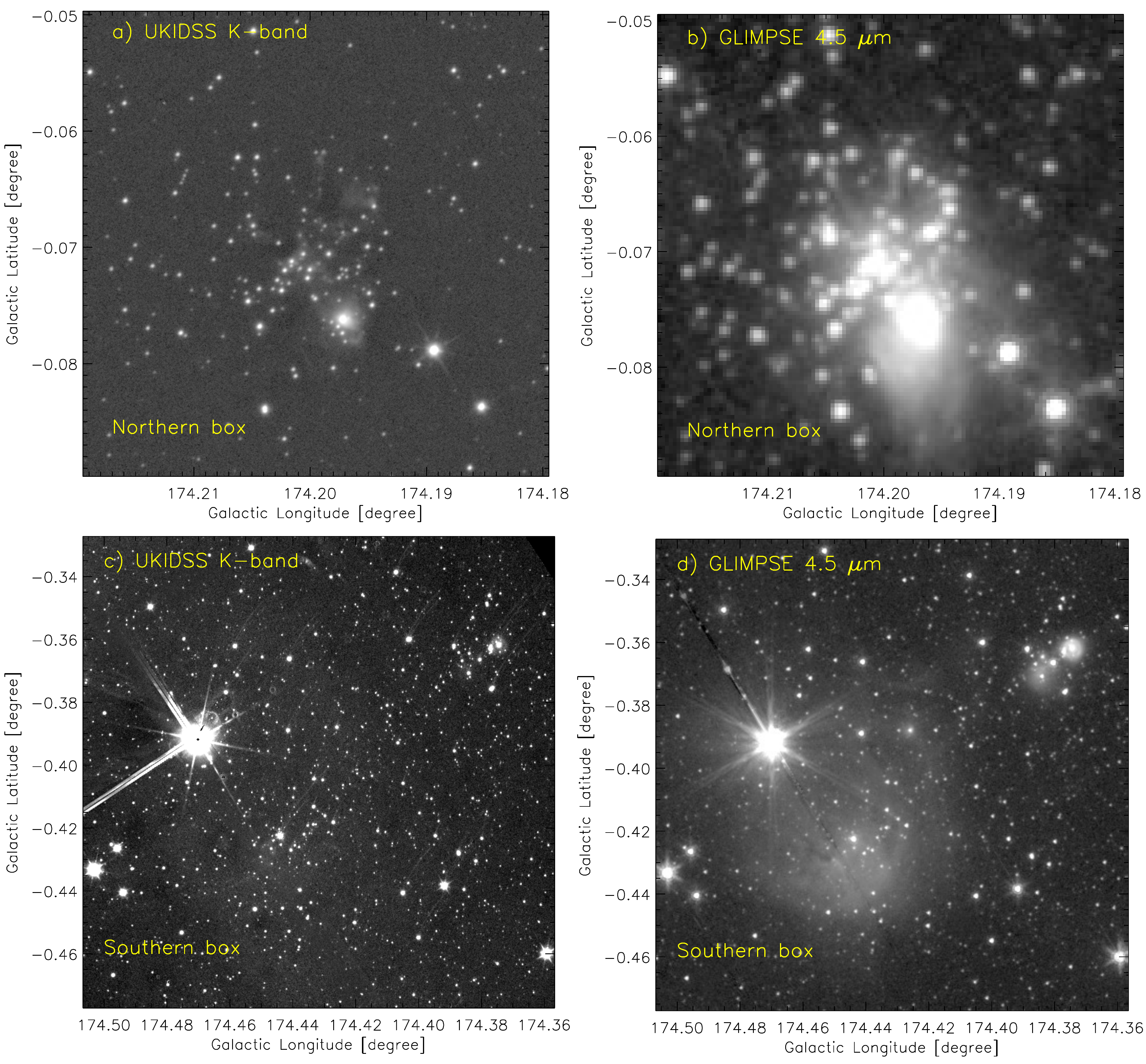}
\caption{a) The panel displays the UKIDSS K-band image of an area highlighted by a solid box (in black) in Figure~\ref{sg1}a (i.e., Northern box). 
b) The GLIMPSE 4.5 $\mu$m image of the area in the direction of Northern box (see a solid box (in black) in Figure~\ref{sg1}a). 
c) The panel displays the UKIDSS K-band image of an area highlighted by a broken box (in yellow) in Figure~\ref{sg1}a (i.e., Southern box). 
d) The GLIMPSE 4.5 $\mu$m image of the area in the direction of Southern box (see a broken box (in yellow) in Figure~\ref{sg1}a).}
\label{sg2}
\end{figure*}

With the help of the mid-infrared emission observed in the {\it WISE} 12 $\mu$m image, a wide-scale environment of AFGL 5142 (area $\sim$25 pc $\times$ $\sim$20 pc) is presented in Figure~\ref{sg1}a. 
The {\it WISE} image is also superimposed with the NVSS 1.4 GHz emission to trace the ionized gas.
An inverted gray-scale image at {\it Herschel} 250 $\mu$m (resolution $\sim$12$''$) is shown in Figure~\ref{sg1}b.
At least two elongated filamentary features ($>$ 10 pc) are visually seen in the {\it Herschel} image, and are highlighted by arrows. 
Two extended and bright regions traced in the image at 12 $\mu$m are found toward the ends of the filaments (see ``northern" and ``southern" boxes in Figure~\ref{sg1}a), which are also spatially extended in the sub-mm image at 250 $\mu$m. 
Interestingly, the star-forming region AFGL 5142 is located at the ``northern end" of the filaments. 
Figures~\ref{sg2}a and~\ref{sg2}b show a zoomed-in view of the ``northern end" using the GPS K-band and the {\it Spitzer} 4.5 $\mu$m images, respectively, revealing the presence of an embedded group of sources. In Figures~\ref{sg2}c and~\ref{sg2}d, the ``southern end" is displayed using the GPS K-band and the {\it Spitzer} 4.5 $\mu$m images, respectively, which also show several embedded point-like sources. 
A quantitative analysis of point-like sources is presented in Section~\ref{subsec:phot1}.
No radio continuum emission at NVSS 1.4 GHz (1$\sigma$ $\sim$0.45 mJy beam$^{-1}$) is observed toward both the ``northern" and ``southern" ends. 

\begin{figure*}
\epsscale{0.53}
\plotone{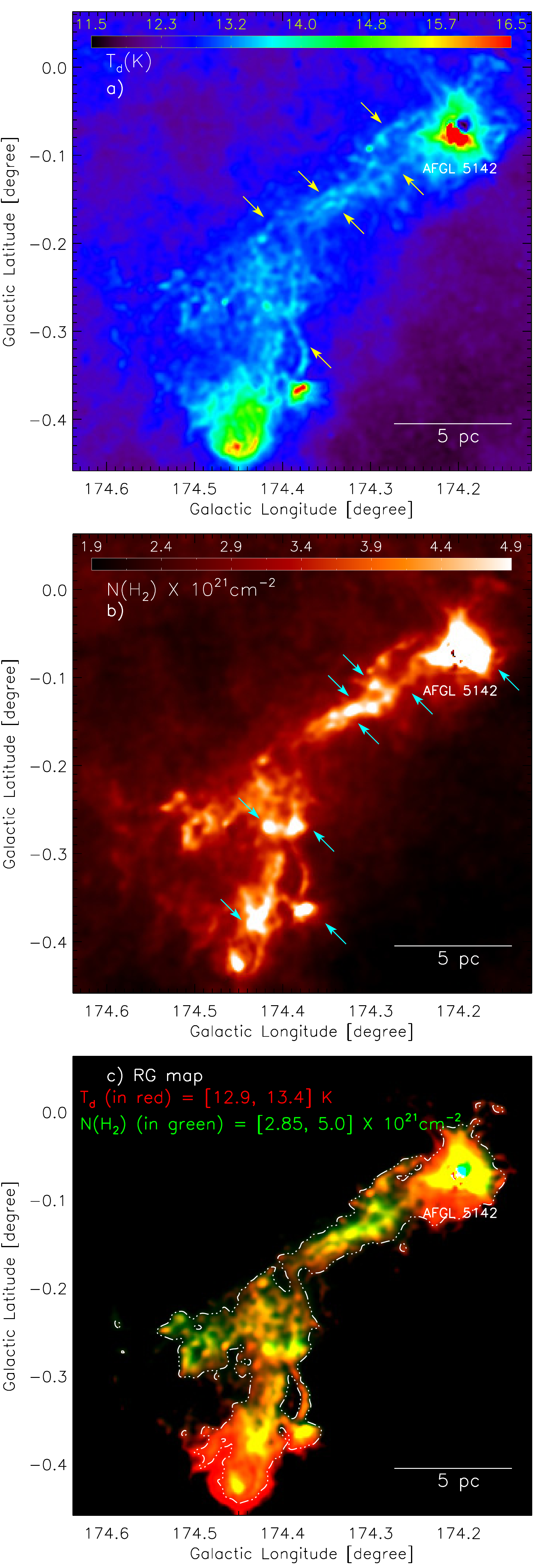}
\caption{{\it Herschel} temperature map (a) and column density ($N(\mathrm H_2)$) map (b) of an area highlighted by a dotted-dashed box (in white) in Figure~\ref{sg1}a. c) Two color-composite image (temperature map (in red) + column density map (in green)). The column density contour (in white) is also overlaid on the map 
with a level of 3.1 $\times$ 10$^{21}$ cm$^{-2}$, indicating an inverted Y-like structure. Arrows in panel ``a" show at least two elongated filaments, 
while arrows in panel ``b" indicate the possible common zones of the filaments.} 
\label{sg3}
\end{figure*}

The {\it Herschel} temperature and column density ($N(\mathrm H_2)$) maps (resolution $\sim$12$''$) are displayed in Figures~\ref{sg3}a and~\ref{sg3}b, respectively. 
These maps were produced as a part of the EU-funded ViaLactea project \citep{molinari10b}, and were obtained from the publicly available site\footnote[3]{http://www.astro.cardiff.ac.uk/research/ViaLactea/}.
In order to generate these maps, \citet{marsh17} utilized the Bayesian {\it PPMAP} procedure \citep{marsh15} to the {\it Herschel} continuum data, and also adopted the optically thin emission assumption in the procedure \citep[see][for more details]{marsh17}. 
Both these {\it Herschel} maps also reveal the elongated filaments as highlighted in Figure~\ref{sg1}b, allowing to infer their physical parameters.  
These filaments also overlap each other at many locations (see arrows in Figure~\ref{sg3}b). Figure~\ref{sg3}c shows a two color-composite image (temperature map (red) and column density map (green)). The column density contour at 3.1 $\times$ 10$^{21}$ cm$^{-2}$ is also overlaid on the color-composite image, revealing an 
inverted Y-like structure. The temperature map traces the filaments in a temperature range of about 12.5--13.5~K. 
We have determined the total mass of the materials distributed within the $N(\mathrm H_2)$ contour of 3.1 $\times$ 10$^{21}$ cm$^{-2}$ (or the inverted Y-like structure), which comes out to be $\sim$6280 M$_{\odot}$. The total mass is computed using the expression, $M_{area} = \mu_{H_2} m_H Area_{pix} \Sigma N(H_2)$, where $\mu_{H_2}$ is the mean molecular weight per hydrogen molecule (i.e., 2.8), $Area_{pix}$ is the area subtended by one pixel (i.e., 6$''$/pixel), and 
$\Sigma N(\mathrm H_2)$ is the total column density \citep[see also][]{dewangan17a}. 

Together, the analysis of the sub-mm data reveals the existence of two elongated filaments in a wide-scale environment of AFGL 5142 and 
several overlapping zones of these filaments (having $N(\mathrm H_2)$ $>$ 4.5 $\times$ 10$^{21}$ cm$^{-2}$).
\subsection{Kinematics of molecular gas}
\label{sec:coem} 
\begin{figure*}
\epsscale{1}
\plotone{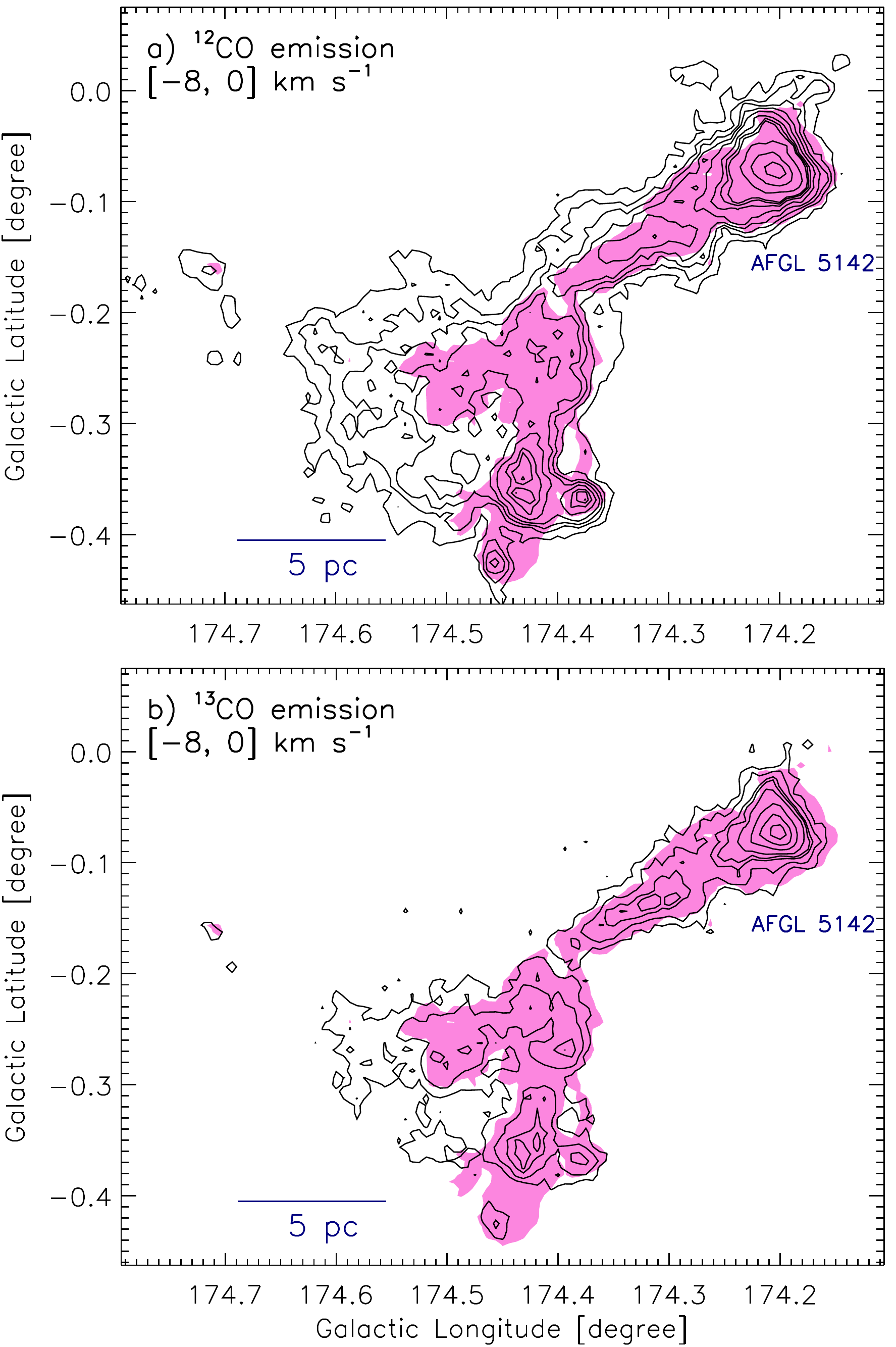}
\caption{a) Integrated $^{12}$CO(J =1$-$0) emission contours at [$-$8, 0] km s$^{-1}$ of an area presented in Figure~\ref{sg1}a. 
The contours are shown with the levels of 107.26 K km s$^{-1}$ $\times$ (0.06, 0.11, 0.16, 0.2, 0.26, 0.33, 0.37, 
0.39, 0.51, 0.7, and 0.88), where 1$\sigma$ $\sim$1.6 K km s$^{-1}$. 
b) Integrated $^{13}$CO(J =1$-$0) emission contours at [$-$8, 0] km s$^{-1}$. The contours are shown with the levels of 38.23 K km s$^{-1}$ $\times$ (0.06, 0.11, 0.2, 0.24, 0.31, 0.47, 0.65, and 0.86), where 1$\sigma$ $\sim$0.7 K km s$^{-1}$.. 
In each panel, the background filled area (in orchid color) shows the inverted Y-like structure, which is observed in the {\it Herschel} 
column density map with a level of 3.1 $\times$ 10$^{21}$ cm$^{-2}$ (see also Figure~\ref{sg3}c).}
\label{sg4}
\end{figure*}

Figures~\ref{sg4}a and~\ref{sg4}b display the distribution of $^{12}$CO (J=1--0) and $^{13}$CO (J=1--0) gas toward the selected target area around AFGL 5142, respectively. The molecular cloud associated with AFGL 5142 is studied in a velocity range of $-$8 to 0 km s$^{-1}$.
The {\it Herschel} column density contour at 3.1 $\times$ 10$^{21}$ cm$^{-2}$ is also overlaid on the integrated maps. 
In general, the $^{12}$CO (1--0) line is typically optically thick compared to the $^{13}$CO (1--0) line. 
Hence, the $^{12}$CO emission can be used to trace the boundary of molecular cloud, while 
the $^{13}$CO emission (n(H$_{2}$) $>$ 10$^{3}$ cm$^{-3}$) allows us to investigate the denser and opaque regions in molecular clouds. 
As per the expectation, the cloud associated with AFGL 5142 appears more extended in 
the $^{12}$CO compared to the $^{13}$CO. 
The $^{13}$CO map also shows the inverted Y-like morphology, which is in agreement with the result derived using the {\it Herschel} column density map.

\begin{figure*}
\epsscale{0.97}
\plotone{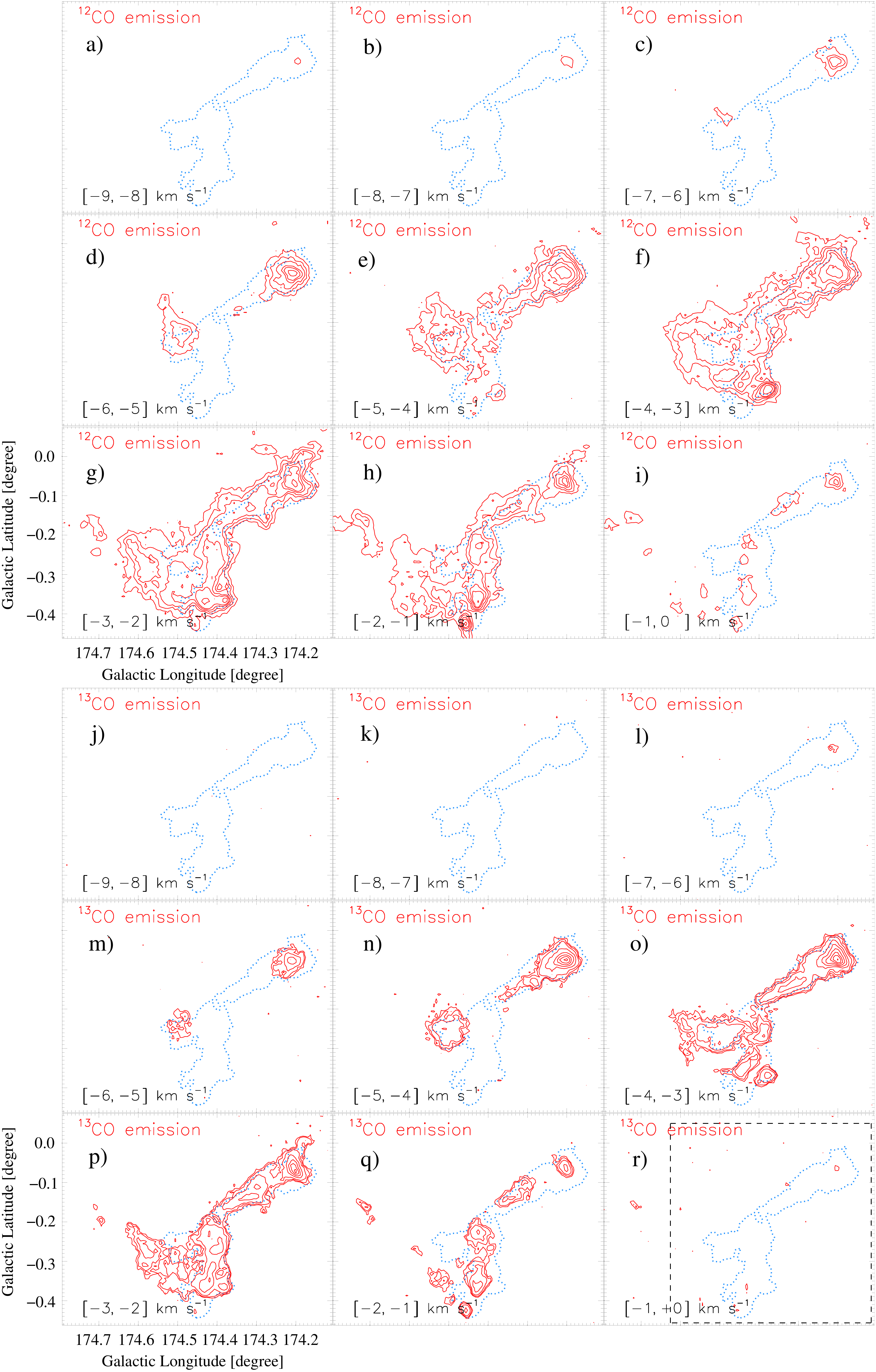}
\caption{a-i) Velocity channel contours of $^{12}$CO emission of an area presented in Figure~\ref{sg1}a. 
j-r) Velocity channel contours of $^{13}$CO emission. 
The molecular emission is integrated over a velocity interval, which is labeled in each panel (in km s$^{-1}$). 
The contour levels of $^{12}$CO are presented with the levels of 2, 5, 8, 12, 15, 20, and 30 K km s$^{-1}$, 
while the $^{13}$CO contours are 0.8, 1.2, 2, 4, 6, 8, 10, 12, 15, 20, and 25 K km s$^{-1}$. 
In each panel, the inverted Y-like structure is indicated by the column density broken contour with a level of 3.1 $\times$ 10$^{21}$ cm$^{-2}$ (see also Figure~\ref{sg3}c).} 
\label{sg5}
\end{figure*}

\begin{figure*}
\epsscale{1.1}
\plotone{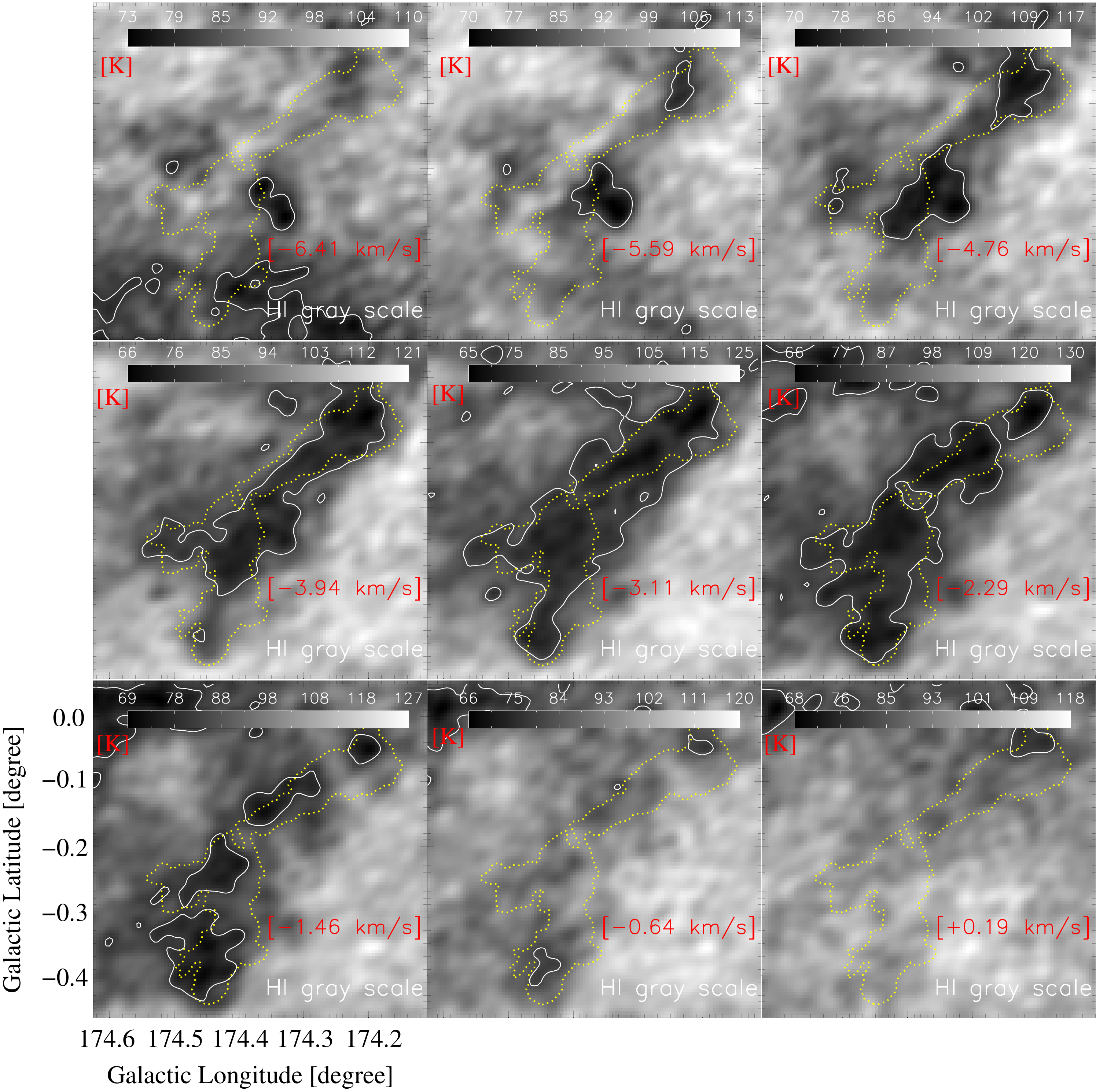}
\caption{The CGPS 21 cm H\,{\sc i} velocity channel maps of the AFGL 5142 site. 
The panel displays an area highlighted by a broken box (in black) in Figure~\ref{sg5}r. 
The velocity value (in km s$^{-1}$) is labeled in each panel. In each panel, the 
inverted Y-like structure is indicated by the column density broken contour (in yellow) with a level of 3.1 $\times$ 10$^{21}$ cm$^{-2}$ (see also Figure~\ref{sg3}c). 
In all the panels, the solid contour (in white) is also shown with the level of 81 K.} 
\label{sg6}
\end{figure*}

In order to further study the distribution of gas, we have examined the velocity channel maps of $^{12}$CO, $^{13}$CO, and H\,{\sc i}. 
In Figure~\ref{sg5}, 
we present $^{12}$CO velocity channel maps within the velocity range from $-$9 to 0 km s$^{-1}$ in steps of 1 km s$^{-1}$ (see panels ``a--i"). The velocity channel maps of $^{13}$CO are also presented in the panels ``j--r" (see Figure~\ref{sg5}). The {\it Herschel} column density contour at 3.1 $\times$ 10$^{21}$ cm$^{-2}$ is also overlaid on each velocity channel map. 
With help of these maps, one can trace two velocity components toward our 
selected field (see maps at [$-2$, $-1$] and [$-5$, $-4$] km s$^{-1}$). 
In Figure~\ref{sg6}, we show the 21 cm H\,{\sc i} velocity channel maps of our selected field. The {\it Herschel} column density contour at 3.1 $\times$ 10$^{21}$ cm$^{-2}$ is also overplotted on each H\,{\sc i} map. 
In the H\,{\sc i} maps, black or dark gray regions are found toward our selected target field, indicating the presence of the absorption features \citep[e.g.,][]{kerton05}. These features are also referred to as the H\,{\sc i} self-absorption (HISA) features \citep[e.g.,][]{kerton05}, which indicate the residual amounts of very cold H\,{\sc i} gas in molecular clouds 
\citep{burton78,baker79,burton81,liszt81}. 
The H\,{\sc i} line data also show the existence of two different cloud components (see maps at $-$2.29 and $-$4.76 km s$^{-1}$). 

\begin{figure*}
\epsscale{0.9}
\plotone{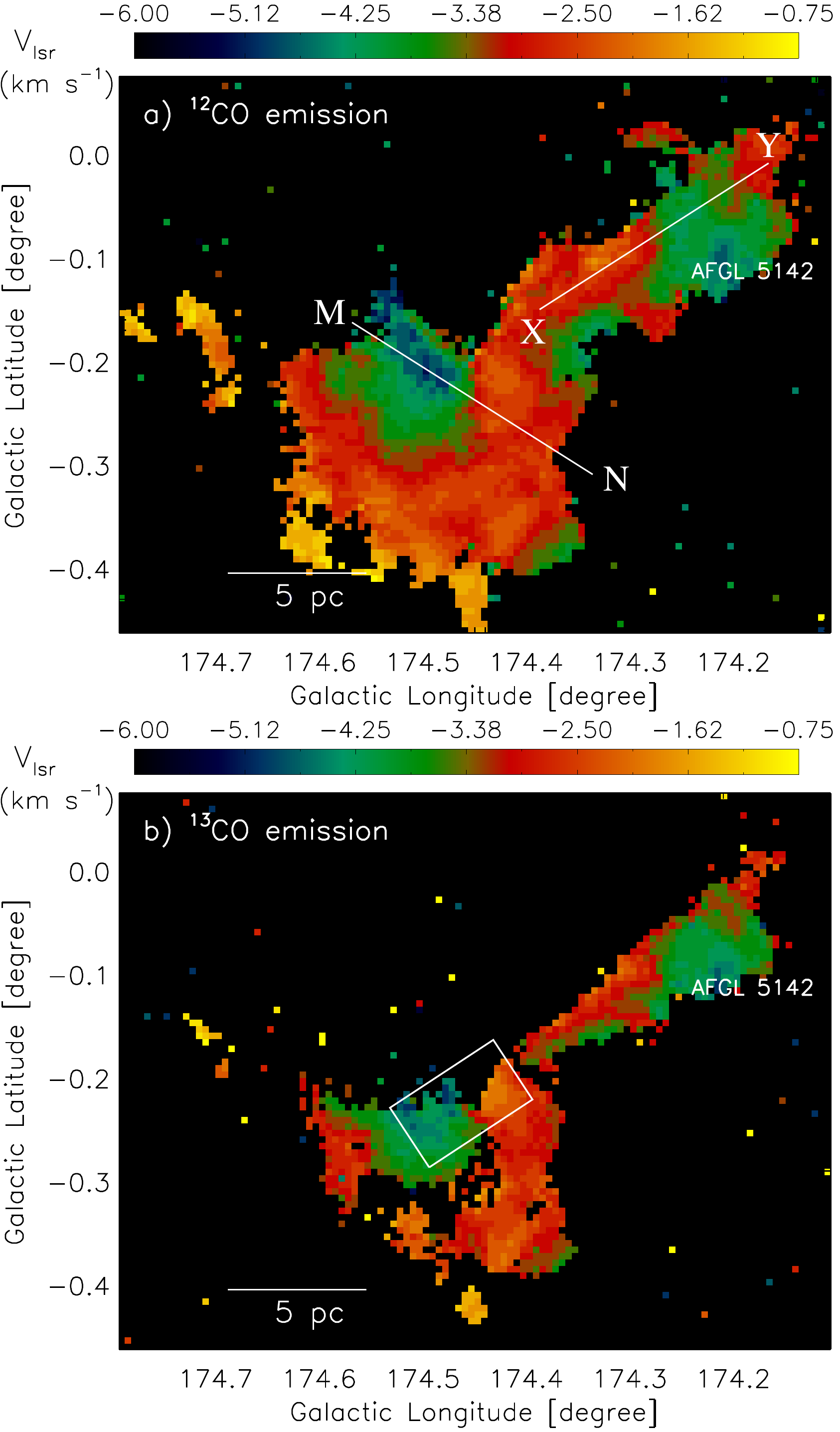}
\caption{a) $^{12}$CO first moment map. 
Two solid lines ``X--Y" and ``M--N" are marked in the figure, where the position-velocity maps are extracted (see Figures~\ref{lsg7}a and~\ref{lsg7}b).
b) $^{13}$CO first moment map. The area highlighted by a solid box (in white) is used for extracting 
the spectra (see Figures~\ref{lsg7}c and~\ref{lsg7}d).}
\label{sg7}
\end{figure*}

\begin{figure*}
\epsscale{0.75}
\plotone{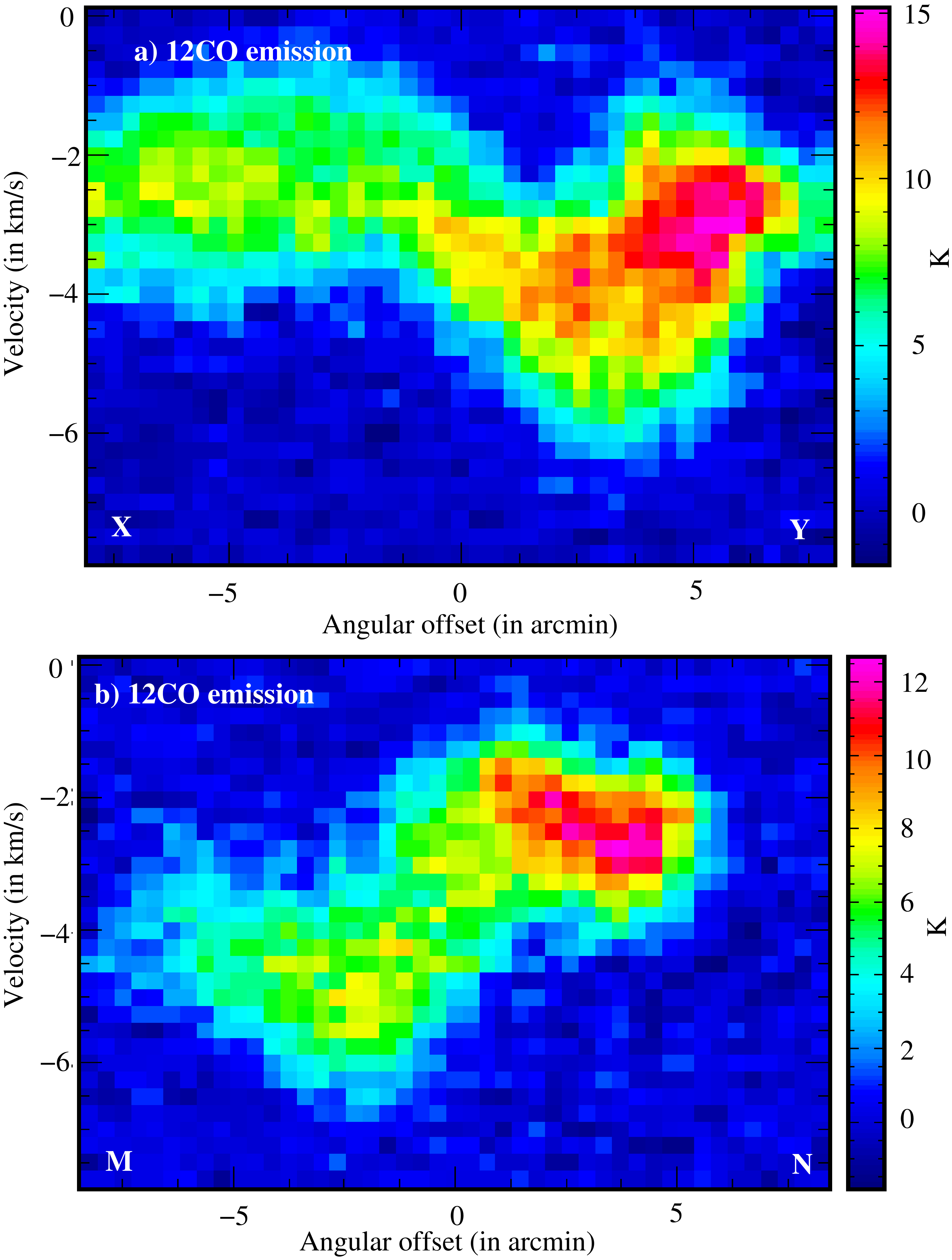}
\epsscale{0.95}
\plotone{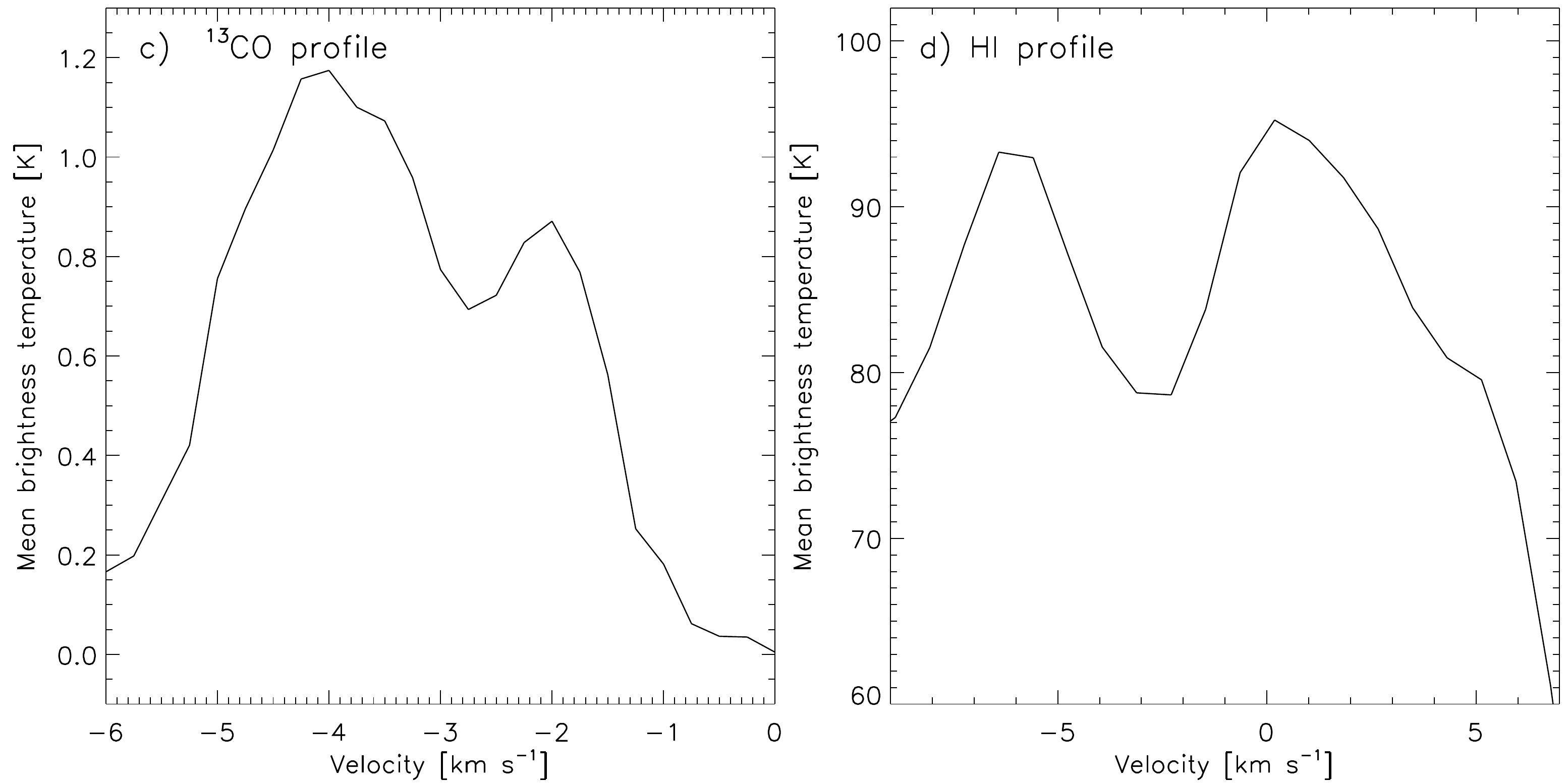}
\caption{a) Position-velocity map of $^{12}$CO along the axis (i.e., ``X--Y") as shown in Figure~\ref{sg7}a (see a solid line in Figure~\ref{sg7}a).
b) Position-velocity map of $^{12}$CO along the axis (i.e., ``M--N") as shown in Figure~\ref{sg7}a (see a solid line in Figure~\ref{sg7}a).
c) The $^{13}$CO profile. d) The H\,{\sc i} profile.
In the panels ``c" and ``d", both the profiles are extracted in the direction of an area marked by a solid box in Figure~\ref{sg7}b. 
These profiles are obtained by averaging the area shown in Figure~\ref{sg7}b.}
\label{lsg7}
\end{figure*}

In Figures~\ref{sg7}a and~\ref{sg7}b, we display the first moment maps of $^{12}$CO and $^{13}$CO, respectively. 
The moment map traces the intensity-weighted mean velocity of the emitting gas. 
The observed velocity structures in these maps suggest two possibilities. In one of the possibilities, there are two distinct filaments in the direction of 
the inverted Y-like structure. Alternatively, the other possibility suggests a single filament with internal velocity gradient toward the inverted Y-like structure. In Figure~\ref{lsg7}a, we present the position-velocity (pv) map of $^{12}$CO in the direction of an axis ``X--Y" (see a solid line in Figure~\ref{sg7}a). 
A velocity gradient is clearly seen toward the Y end of the axis ``X--Y", where AFGL 5142 is located. 
Figure~\ref{lsg7}b displays the pv map of 
$^{12}$CO along an axis ``M--N" (see a solid line in Figure~\ref{sg7}a). In Figure~\ref{lsg7}b, the pv map suggests the velocity connection of two velocity components (around $-$1.5 and $-$4.5 km s$^{-1}$). 
The observed $^{13}$CO and H\,{\sc i} profiles are shown in Figures~\ref{lsg7}c and~\ref{lsg7}d, respectively. 
Both the spectra show two velocity peaks, and are generated by averaging the area highlighted by a solid box in Figure~\ref{sg7}b. 

\begin{figure*}
\epsscale{0.58}
\plotone{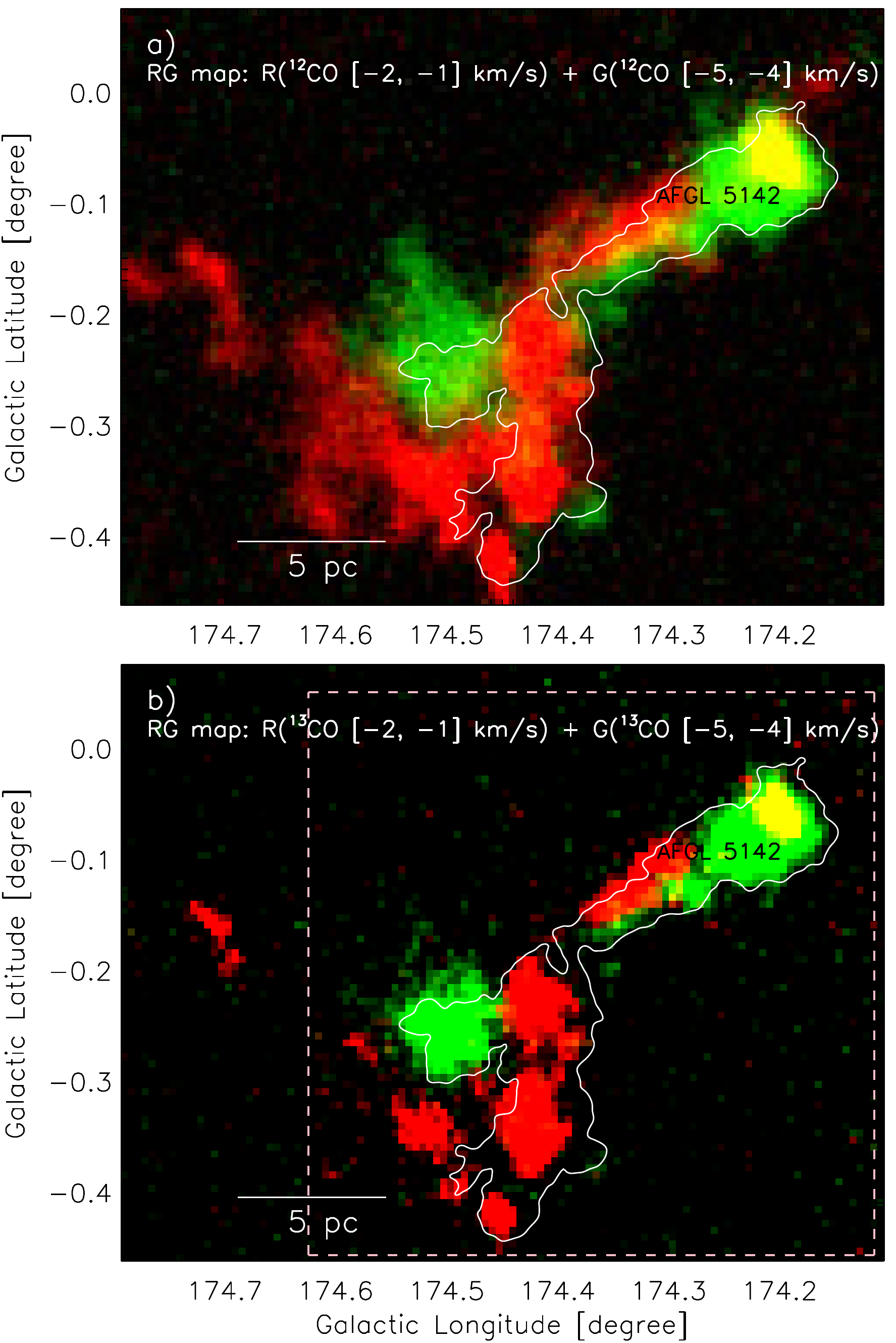}
\epsscale{0.58}
\plotone{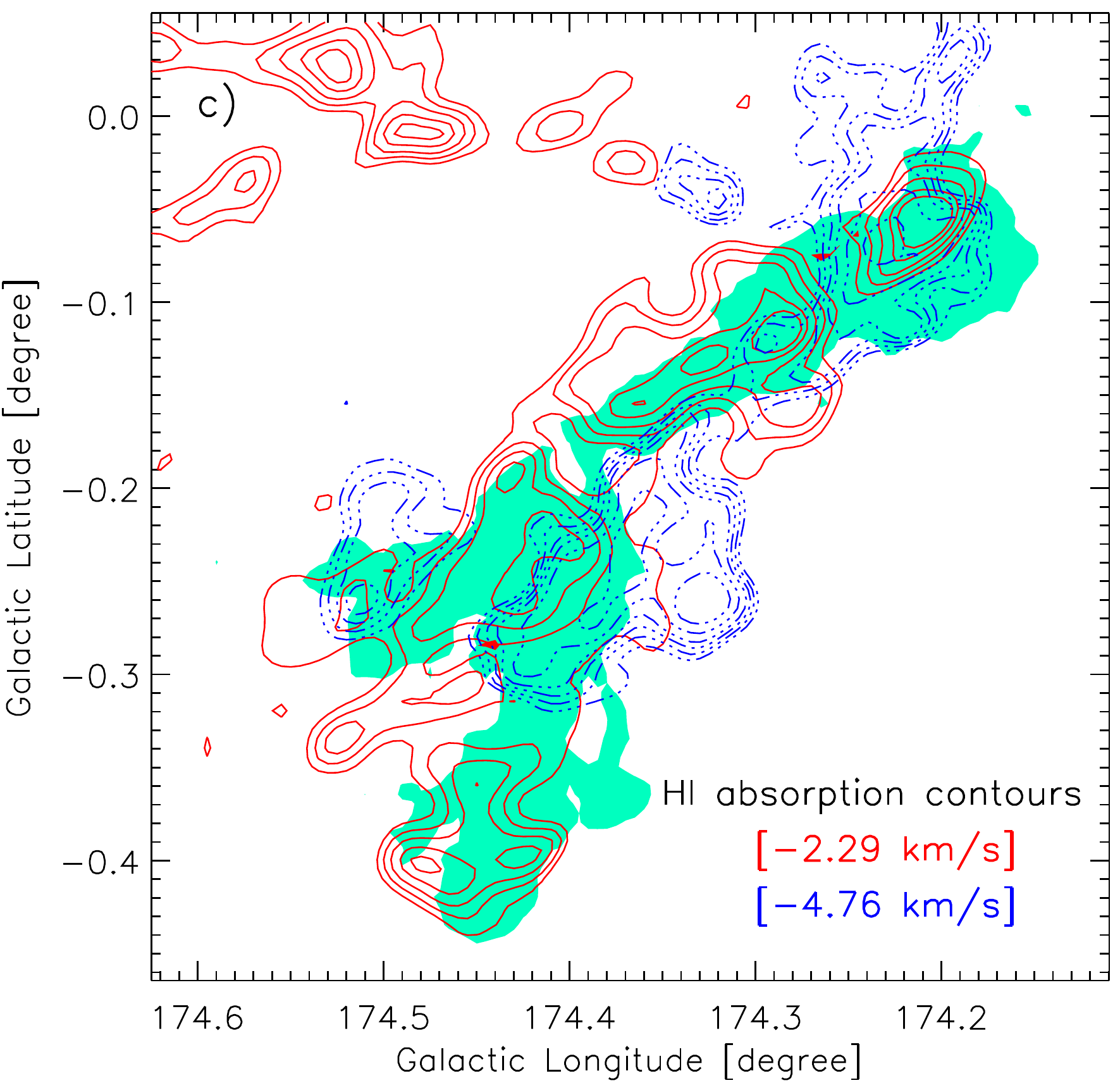}
\caption{a) Two color-composite image of the site AFGL 5142 with the $^{12}$CO maps at [$-$2, $-$1] and [$-$5, $-$4] km s$^{-1}$ in red and green, respectively (see Figures~\ref{sg5}h and~\ref{sg5}e).
b) Two color-composite image made using the $^{13}$CO maps at [$-$2, $-$1] and [$-$5, $-$4] km s$^{-1}$ in red and green, respectively (see Figures~\ref{sg5}q and~\ref{sg5}n).
c) The H\,{\sc i} velocity channel maps (at $-$2.29 and $-$4.76 km s$^{-1}$) of an area highlighted by a broken box in Figure~\ref{sg8}b. 
The HISA features traced in the maps at $-$2.29 and $-$4.76 km s$^{-1}$ are shown by solid red and broken blue contours, respectively (see also Figure~\ref{sg6}).  
The solid contours (in red) are shown with the levels of 70, 72, 75, 78, and 81 K, decreasing from the outer to 
the inner regions. The broken contours (in blue) are presented with the levels of 74, 77, 80, 82, 84, and 86 K, decreasing from the outer to the inner regions. The background filled area (in light green color) shows the inverted Y-like structure, which is observed in the {\it Herschel} column density map with a level of 3.1 $\times$ 10$^{21}$ cm$^{-2}$ (see also Figure~\ref{sg3}c). In the panels ``a" and ``b", the 
inverted Y-like structure is indicated by the column density contour (in white).}
\label{sg8}
\end{figure*}

In Figures~\ref{sg8}a and~\ref{sg8}b, we show color-composite images of the site AFGL 5142 with the $^{12}$CO and $^{13}$CO maps 
at [$-2$, $-1$] and [$-5$, $-4$] km s$^{-1}$ in red and green, respectively. 
The inverted Y-like structure observed in the {\it Herschel} column density map is also indicated by a solid contour. 
We find that these color-composite maps can reproduce the observed velocity field as traced in the first moment maps of $^{12}$CO and $^{13}$CO. It also favours the existence of two intertwined filamentary clouds along the line of sight. 
Interestingly, the inverted Y-like structure can also be explained by the two elongated clouds (around $-$1.5 and $-$4.5 km s$^{-1}$), which are also spatially overlapped. In other words, a complimentary distribution of molecular gas at [$-2$, $-1$] and [$-5$, $-4$] km s$^{-1}$ produces the observed inverted Y-like structure. Hence, it is unlikely the existence of a single filament with internal velocity gradient toward the inverted Y-like structure. 
Figure~\ref{sg8}c shows the contours of the HISA features traced in the 21 cm H\,{\sc i} velocity channel maps at $-$2.29 and $-$4.76 km s$^{-1}$. All the observed results derived using the molecular line data are also supported by the distribution of the cold H\,{\sc i} gas (see Figures~\ref{sg6} and~\ref{sg8}c).

\begin{figure*}
\epsscale{1.15}
\plotone{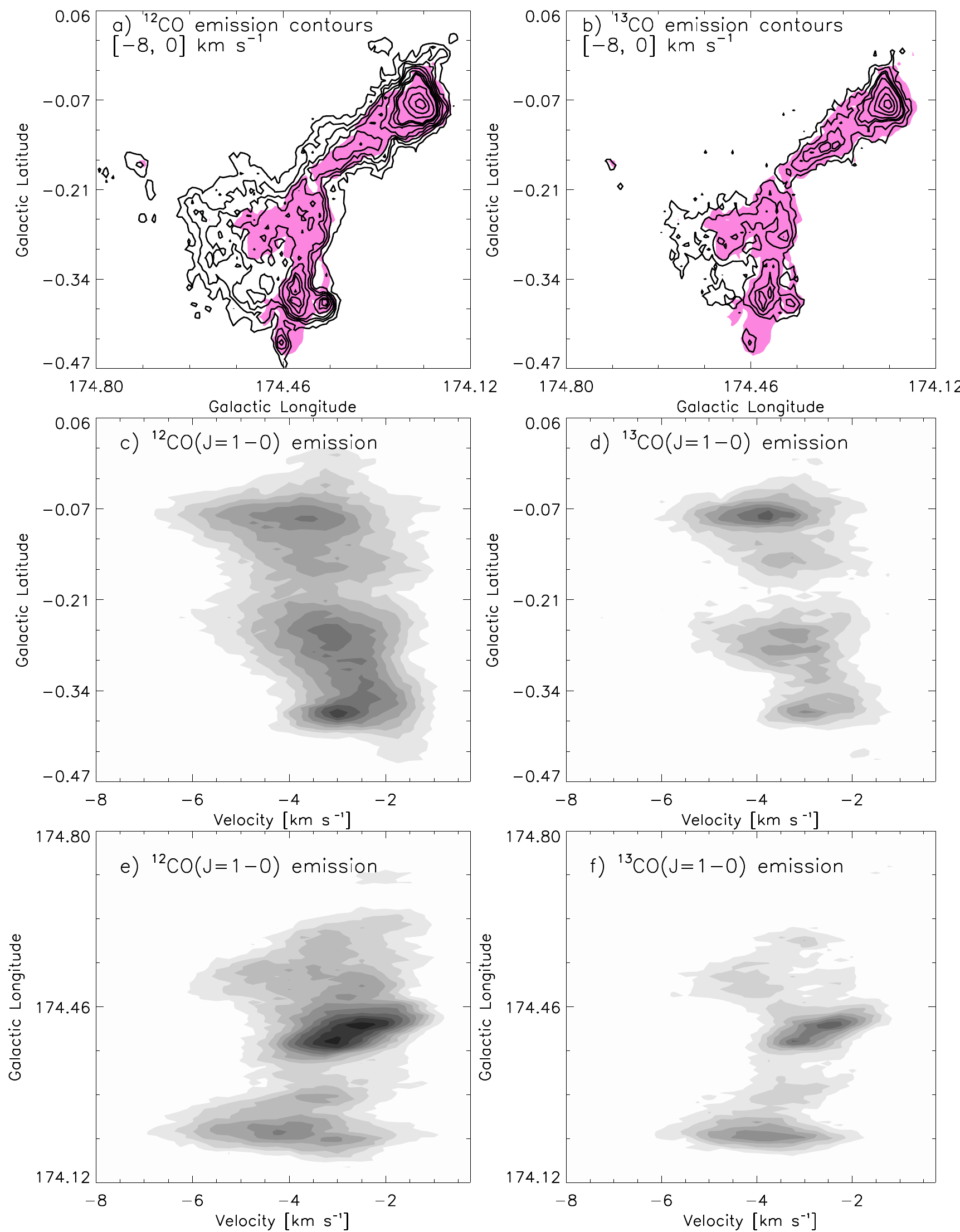}
\caption{a) Integrated $^{12}$CO emission contours at [$-$8, 0] km s$^{-1}$. 
The $^{12}$CO contours and the background filled area (in orchid color) are the same as shown in Figure~\ref{sg4}a. 
b) Integrated $^{13}$CO emission contours at [$-$8, 0] km s$^{-1}$. 
The $^{13}$CO contours and the background filled area (in orchid color) are the same  as shown in Figure~\ref{sg4}b. 
c) Latitude-velocity map of $^{12}$CO. d) Latitude-velocity map of $^{13}$CO.
e) Longitude-velocity map of $^{12}$CO. f) Longitude-velocity map of $^{13}$CO. 
In the Latitude-velocity plots (panels ``c" and ``d"), the molecular emission is integrated over the longitude range from 174$\degr$.12 to 174$\degr$.8.
In the Longitude-velocity plots (panels ``e" and ``f"), the molecular emission is integrated over the latitude range from $-$0$\degr$.47 to 0$\degr$.06.}
\label{sg9}
\end{figure*}

In Figure~\ref{sg9}, we present the integrated $^{12}$CO and $^{13}$CO intensity maps 
and the Galactic pv maps. The integrated maps are similar as shown in Figure~\ref{sg4}. 
In the pv maps, velocity gradients are observed in the direction of the both ends and the center of the cloud.
A detailed discussions on these results are presented in Section~\ref{sec:disc}.
\subsection{Spatial distribution of YSOs in the site AFGL 5142}
\label{subsec:phot1}
\begin{figure*}
\epsscale{1}
\plotone{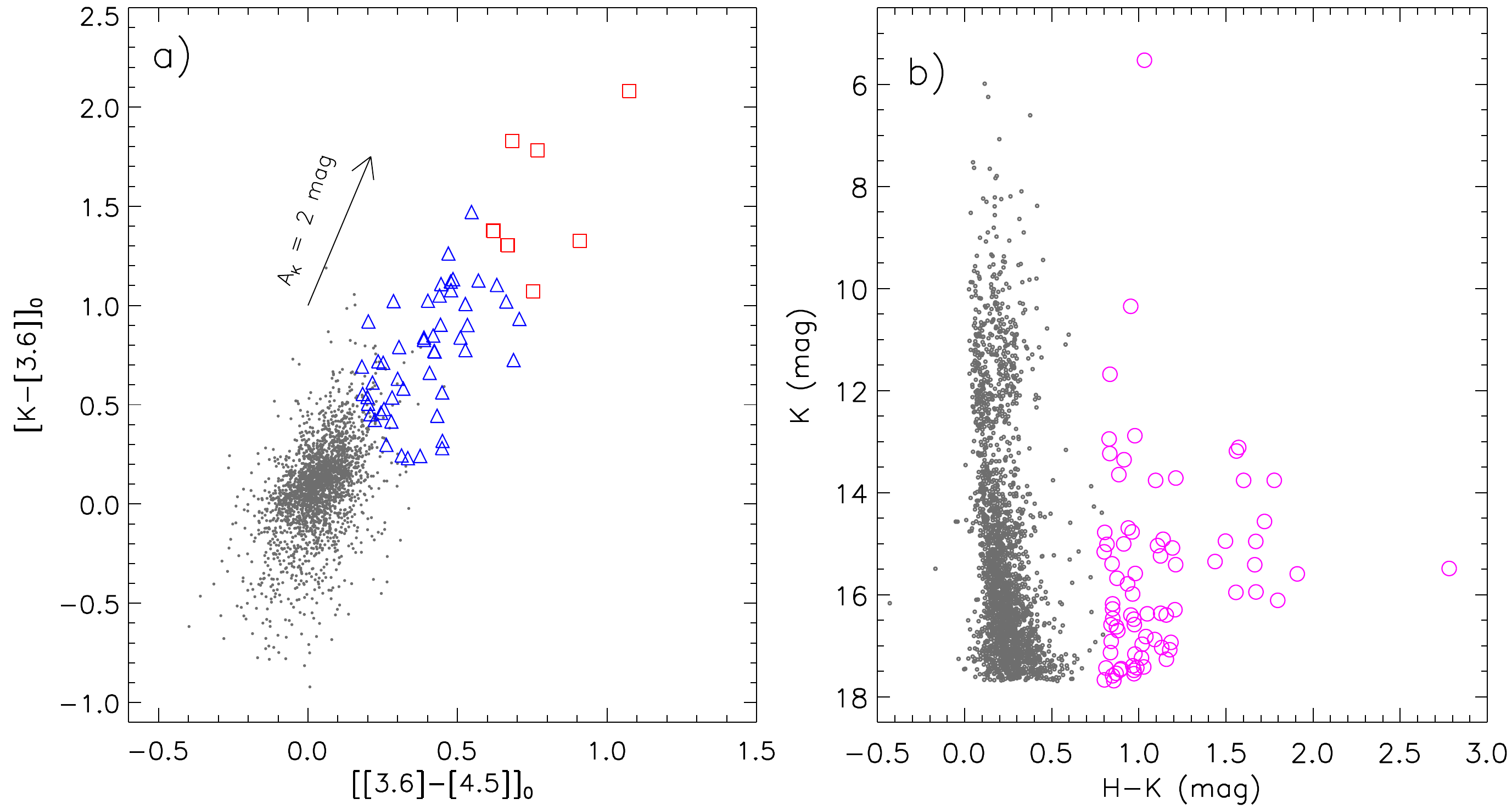}
\epsscale{0.56}
\plotone{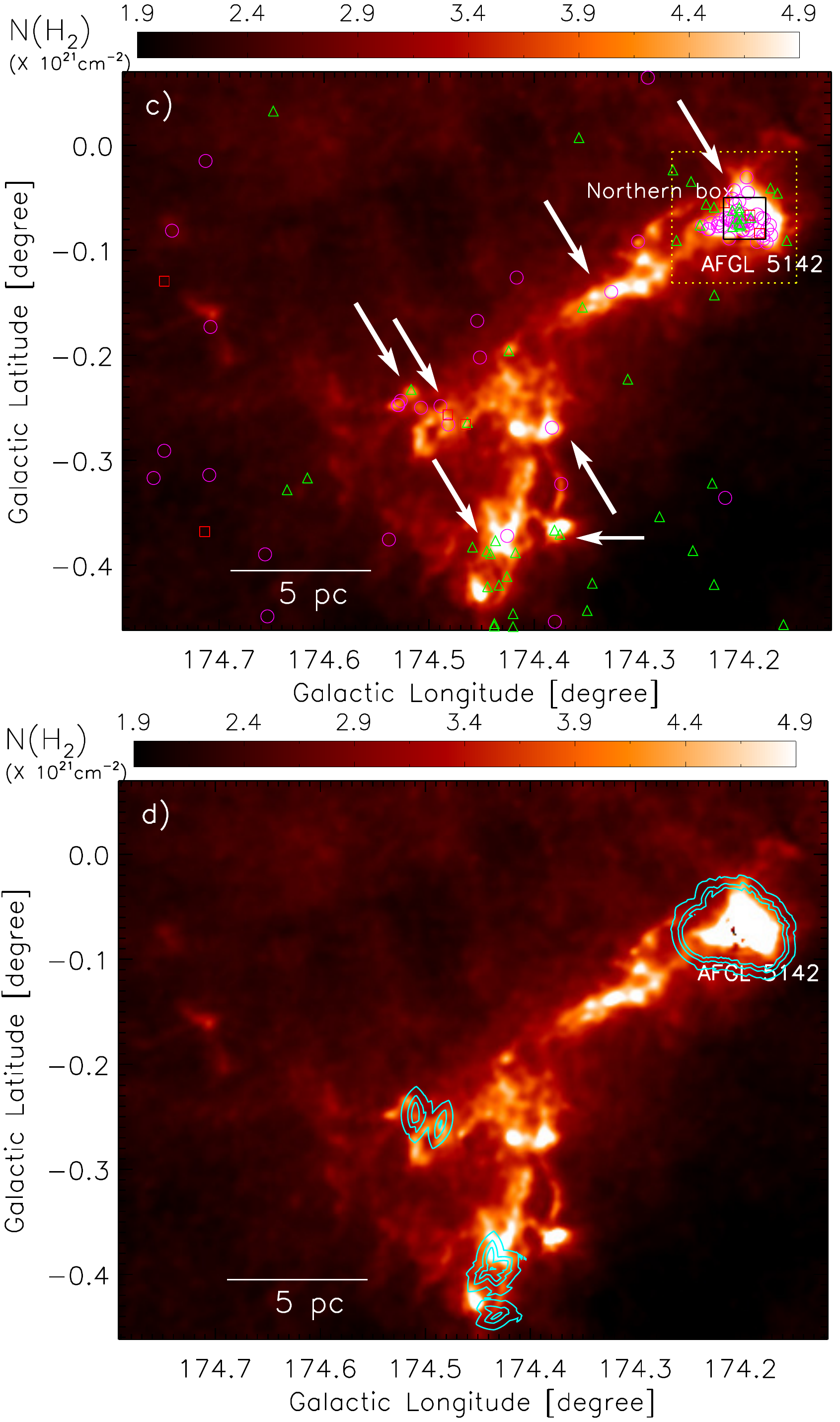}
\caption{Selection of YSOs embedded in the region probed in this paper (see Figure~\ref{sg1}a). 
a) The dereddened color-color plot ([K$-$[3.6]]$_{0}$ $vs$ [[3.6]$-$[4.5]]$_{0}$) of point-like sources.
The extinction vector is drawn using the average extinction laws from \citet{flaherty07}. 
Class~I and Class~II YSOs are marked by red squares and open blue triangles, respectively. 
b) Color-magnitude plot (H$-$K/K) of point-like sources. The objects with color-excess are shown by magenta circles. 
In the panels ``a" and ``b", the dots in gray color refer the stars with only photospheric emission. 
c) Overlay of all the selected YSOs on the {\it Herschel} column density map. The same symbols are used here as adopted in Figures~\ref{sg10}a and~\ref{sg10}b. Arrows indicate the presence of YSOs in the direction of the common zones of the elongated filaments. d) Overlay of surface density contours of YSOs (in cyan) on the {\it Herschel} column density map. The contours are plotted with the levels of 1, 1.5, and 2 YSOs pc$^{-2}$.}
\label{sg10}
\end{figure*}

This section deals with the selection of infrared-excess sources embedded in the site AFGL 5142.
In this connection, we have employed two selection schemes using the GLIMPSE360, UKIDSS-GPS, and 2MASS photometric data, 
which are the dereddened color-color (i.e. [K$-$[3.6]]$_{0}$ and [[3.6]$-$[4.5]]$_{0}$) and 
the NIR color-magnitude (i.e. H$-$K/K) schemes. 
One can find more detailed descriptions of these schemes in \citet{dewangan17a} \citep[see also][]{gutermuth09}.
In Figures~\ref{sg10}a and~\ref{sg10}b, we show the dereddened color-color plot 
(i.e. [K$-$[3.6]]$_{0}$ and [[3.6]$-$[4.5]]$_{0}$) and the NIR color-magnitude plot 
(i.e. H$-$K/K) of point-like objects, respectively. 
In the dereddened color-color scheme, we computed the dereddened colors using the color excess ratios listed 
in \citet{flaherty07}, and separated YSOs against the possible dim extragalactic contaminants. 
This scheme gives 57 (7 Class~I and 50 Class~II) YSOs (see Figure~\ref{sg10}a). 
In the NIR color-magnitude scheme, we have utilized a color criterion (i.e., H$-$K $>$ 0.8 mag) to select infrared excess sources, which is determined based on the color-magnitude analysis of the nearby control field.
We identify 76 additional YSO candidates using this scheme in our selected region (see Figure~\ref{sg10}b).
Using these classification schemes lead to select a total of 133 YSOs in our target field. 
In Figure~\ref{sg10}c, the positions of all the selected YSOs are overlaid on the {\it Herschel} column density map. 
We find that the embedded YSOs are located to those points where filaments overlap (including the ends of filaments) (see thick arrows in Figure~\ref{sg10}b). 

\begin{figure*}
\epsscale{0.8}
\plotone{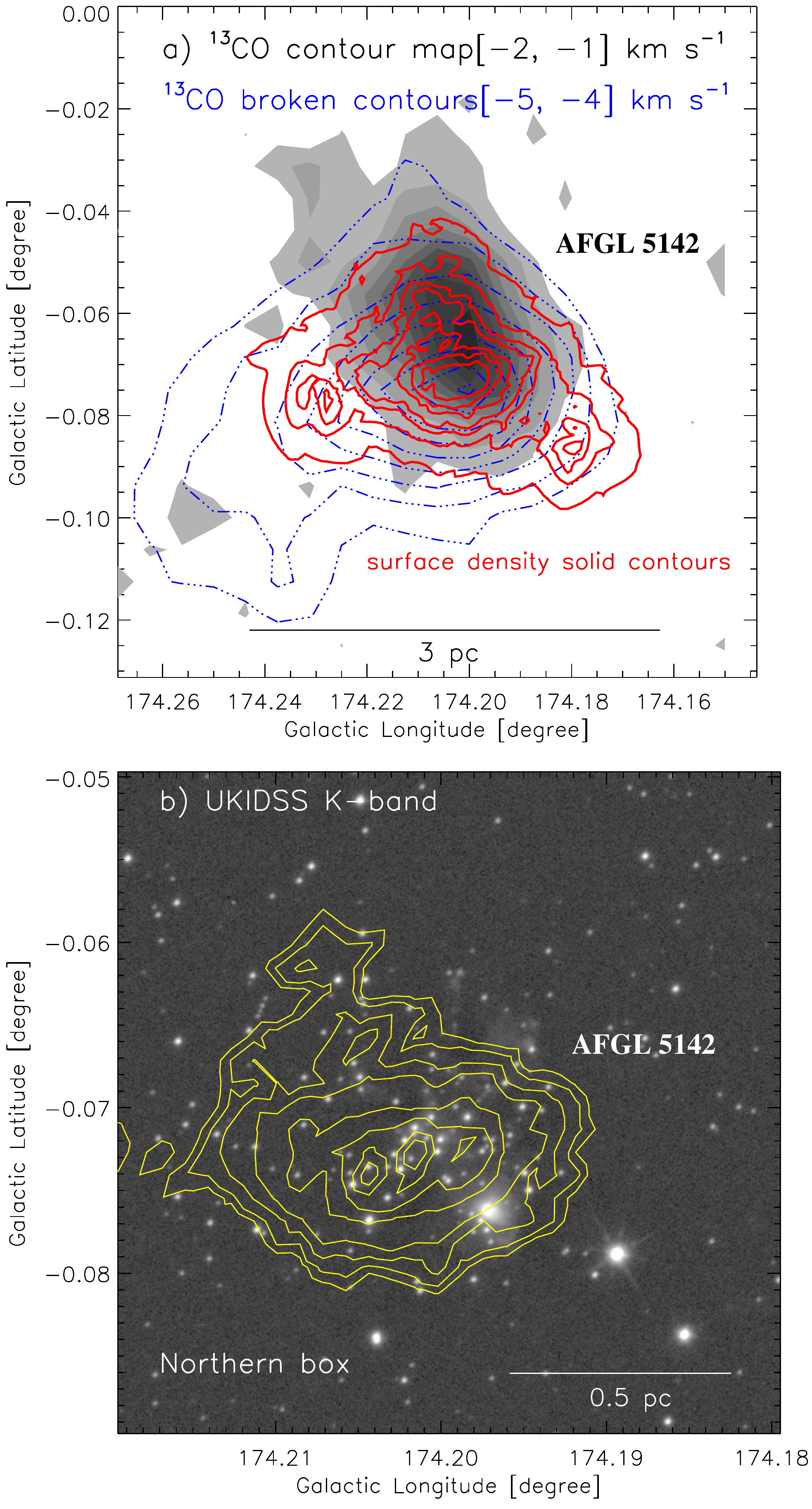}
\caption{a) Overlay of surface density contours of YSOs (in red) on the $^{13}$CO maps at [$-2$, $-1$] 
and [$-5$, $-4$] km s$^{-1}$ of an area located in the Galactic northern direction (see a dashed box (in yellow) in Figure~\ref{sg10}c). The filled contours at [$-2$, $-1$] km s$^{-1}$ are shown with the levels of 0.48, 0.95, 1.43, 1.91, 2.38, 2.86, 3.34, 3.82, 4.29, and 4.68 K km s$^{-1}$, 
while the blue dotted-dashed contours at [$-5$, $-4$] km s$^{-1}$ are plotted with the levels of 2.54, 3.82, 5.1, 6.36, 7.63, 8.91, 10.18, 11.45, and 12.47 K km s$^{-1}$. The surface density contours (in red) are plotted with the levels of 5, 10, 15, 20, 35, 50, and 80 YSOs pc$^{-2}$. 
b) Overlay of surface density contours (in yellow) of YSOs on the UKIDSS K-band image toward AFGL 5142 (see a solid box (in black) in Figure~\ref{sg10}c). The contours are plotted with the levels of 28, 32, 40, 50, 80, 120, 180, and 240 YSOs pc$^{-2}$.}
\label{sg11}
\end{figure*}

Using the nearest-neighbour (NN) technique \citep[e.g.,][]{gutermuth09} and following the procedures described in \citet{dewangan15}, the surface density map of YSOs is also computed in this paper \citep[also see equation in][]{dewangan15}. 
Figure~\ref{sg10}d shows the surface density contours of YSOs overlaid on the {\it Herschel} column density map, indicating the presence of groups of YSOs in the northern and southern parts of the inverted Y-like structure. 
The surface density contours are shown with the levels of 1, 1.5, and 2 YSOs pc$^{-2}$. 
In Figure~\ref{sg11}a, a zoomed-in view of the northern part of the inverted Y-like structure is displayed using the $^{13}$CO maps at [$-2$, $-1$] and [$-5$, $-4$] km s$^{-1}$, which hosts AFGL 5142. 
We have also overlaid the surface density contours of YSOs on the $^{13}$CO maps, and the contours are 
plotted with the levels of 5, 10, 15, 20, 35, 50, and 80 YSOs pc$^{-2}$. 
In the direction of AFGL 5142, a cluster of YSOs is found to be spatially extended within a scale of $\sim$3 pc, and are located toward the common zone of two clouds around 1.5 and 4.5 km s$^{-1}$. Based on the surface density value, the utility of the identification 
of a cluster of sources is discussed in \citet{bressert10} (see their paper for more details). Figure~\ref{sg11}b further displays the surface density contours of YSOs superimposed on the UKIDSS K-band image toward AFGL 5142.
The surface density contours are shown with the levels of 28, 32, 40, 50, 80, 120, 180, and 240 YSOs pc$^{-2}$.
It implies that SF activities are very intense toward the northern end of the inverted Y-like structure (i.e., AFGL 5142). 
Furthermore, at least two peaks of the surface density with the values of 170--240 YSOs pc$^{-2}$ are also observed 
within a scale of 0.3 pc (see Figure~\ref{sg11}b), and are seen toward the SMA 1.1 mm continuum cores \citep[see Figure~3 in][]{liu16}. 
\section{Discussion}
\label{sec:disc}
Earlier, several signposts of SF (i.e., outflows, 22-GHz water maser, and 6.7-GHz methanol maser) have been reported toward AFGL 5142 \citep[see][and references therein]{liu16}. 
Based on the detection of the 6.7-GHz methanol maser and the radio continuum emission, AFGL 5142 is considered as an active site of massive SF \citep[see][and references therein]{goddi11,liu16}. 
Using the HCO$^{+}$(1--0) spectrum, two velocity peaks (around $-$1.5 and $-$4.5 km s$^{-1}$) have been observed toward AFGL 5142 
\citep{hunter99,liu16}. Using the H$^{13}$CN profile, \citet{colzi18} also found two peaks toward AFGL 5142-MM in a velocity range of [$-$10, 0] km s$^{-1}$. \citet{liu16} reported a hierarchical network of filaments using the NH$_{3}$ (1, 1) line data. 
With the help of the pv diagrams of NH$_{3}$ (1, 1), the existence of molecular gas in a velocity range of [$-$5.5, $-$1.5] km s$^{-1}$ has 
been found by \citet{liu16} (see Figure~11 in their paper). They traced one filament in a velocity 
range of [$-$5.5, $-$3.0] km s$^{-1}$, while the other one was observed at [$-$3.0, $-$1.5] km s$^{-1}$ (see Figure~11 in their paper). 
All these results favour the presence of at least two velocity components in the direction of the massive star-forming region AFGL 5142. 
However, in the literature, there is no information available concerning the filaments and different velocity components at a large-scale.

In Section~\ref{subsec:u1}, we have investigated an embedded structure in a wide-scale environment around AFGL 5142, which is the 
inverted Y-like morphology (mass $\sim$6280 M$_{\odot}$; major axis $\sim$19 pc). 
The {\it Herschel} map also enables us to identify at least two elongated filaments (lengths $>$ 10 pc), 
which are embedded within the the inverted Y-like morphology. In parallel to the outcomes derived from the {\it Herschel} data sets, the distribution of the $^{12}$CO, $^{13}$CO, and H\,{\sc i} gas toward the structure is also examined in Section~\ref{sec:coem}. In the direction of the structure, at least two different elongated filamentary molecular clouds (around $-$1.5 and $-$4.5 km s$^{-1}$) are traced using the CO line data, and are also linked spatially as well as in velocity (see Section~\ref{sec:coem}, and also Figure~\ref{lsg7}). The first moment maps of CO show a pair of nearly twisted filamentary clouds at different positions and velocities, 
revealing their spatial overlapping (see Figures~\ref{sg7}a and~\ref{sg7}b). 
To our knowledge, such observational outcome is rare, and is also very promising. 
The overlapping zones include both the ends and the central part of the filamentary clouds, where $N(\mathrm H_2)$ $>$ 4.5 $\times$ 10$^{21}$ cm$^{-2}$. 
Interestingly, the spatial distribution of molecular gas at [$-2$, $-1$] and [$-5$, $-4$] km s$^{-1}$ produces the observed inverted Y-like structure. The analysis of the cold H\,{\sc i} gas also supports these observational outcomes (see Figures~\ref{sg6}a and~\ref{sg8}c). 
All these results enable us to suggest the interaction between filamentary clouds along the line of sight. 
Hence, one can also expect SF activities at the shock-compressed layer of gas due to the collision of filamentary clouds 
\citep[e.g.,][and references therein]{habe92,nakamura14,dewangan18}. Furthermore, the collision of two clouds has been proposed as a possible triggering mechanism of massive SF \citep[e.g.,][]{inoue13,fukui14}. 

In Section~\ref{subsec:phot1}, the spatial distribution of YSOs shows SF activities toward the 
inverted Y-like structure. Noticeable YSOs are found toward the common zones of the elongated clouds. 
Furthermore, very strong SF activities are found toward the northern end of the structure containing AFGL 5142, 
where the clustering of YSOs is spatially extended within a scale of $\sim$3 pc (see Section~\ref{subsec:phot1} and also Figure~\ref{sg11}a). 
In the direction of AFGL 5142, the embedded cluster is identified at the overlapping area of the two clouds around 1.5 and 4.5 km s$^{-1}$ (see Figure~\ref{sg11}a). The observed velocity separation of the two clouds is about $\sim$3 km s$^{-1}$. With the existing data sets, it is not possible to know the exact 
viewing angle of the collision. Hence, the viewing angle of the collision to the line of sight is assumed to be 45$\degr$ in this paper. 
We have found the collision length-scale (l$_{fcs}$) to be $\sim$4.2 pc (= 3.0 pc/sin(45$\degr$)), while 
the observed relative velocity (${v_{\rm rel}}$) is estimated to be $\sim$4.2 km s$^{-1}$ (= 3 km s$^{-1}$/cos(45$\degr$)).
Using the value of l$_{fcs}$ and ${v_{\rm rel}}$, one can compute the time-scale of the accumulation of material at the collision points or the collision time-scale (see equation~2 in \citet{henshaw13} and also \citet{mckee07}), which is given by
\begin{equation}
t_{\rm accum} = 2.0\,\bigg(\frac{\rm l_{fcs}}{0.5\,{\rm pc}} \bigg) \bigg(\frac{v_{\rm
rel}}{5{\rm \,km\,s^{-1}}}\bigg)^{-1}\bigg(\frac{n_{\rm pstc}/n_{\rm
prec}}{10}\bigg)\,{\rm Myr} 
\end{equation}
where, l$_{fcs}$ and ${v_{\rm rel}}$ are defined earlier, n$_{prec}$ is the mean density of pre-collision region, and n$_{pstc}$ is the mean density of post-collision region. In this paper, the exact ratio of the mean densities of the pre- and post-collision regions is unknown. 
Hence, we adopt a range of the ratios of densities (i.e. 0.5--10) then 
a range of the typical collision timescales is computed to be $\sim$1--20 Myr. 
Previously observed 6.7-GHz methanol maser also indicates the presence of early phase of massive SF ($<$ 0.1 Myr) in AFGL 5142 \citep[e.g.,][]{liu16}. Considering the mean ages of Class~I and Class~II YSOs of 
$\sim$0.44 Myr and $\sim$1--3 Myr, respectively \citep{evans09}, 
we find that the collision timescale is old enough to influence the formation 
of the youngest protostars and/or massive star(s) in our selected target field. The dynamical time of the high-velocity CO outflows and the EWBO was also computed to be $\sim$10$^{4}$ yr \citep[e.g.,][]{zhang07,liu16}. The results presented by \citet{liu16} accordingly indicate the existence of the earliest stages of SF in AFGL 5142. Hence, all these observational clues favour the onset of filament-filament collision in our target site, which seems to explain SF history in the site AFGL 5142. This process might have also influenced the formation of massive star(s) in the site AFGL 5142. 
In the literature, we have also found some other star-forming sites (such as, W33A \citep{galvan10}, Serpens \citep{duarte11,nakamura14}, L1641-N \citep{nakamura12}, Rosette Nebula \citep{schneider12}, Infrared dark cloud G035.39$-$00.33 \citep{henshaw13}, and Sh 2-237 \citep{dewangan17a}), 
where the origin of young stellar cluster is explained by the collision/interaction of filaments. 

In a wide-scale area around AFGL 5142, high-resolution molecular line data will be helpful to further examine the collision/interaction process of filaments.
\section{Summary and Conclusions}
\label{sec:conc}
In this paper, to understand the ongoing physical mechanism, we have examined a large-scale environment (area $\sim$25 pc $\times$ $\sim$20 pc) around a star-forming site, AFGL 5142. The present work is carried out using the multi-wavelength data sets spanning radio to NIR wavelengths, which have allowed us to study the gas and dust in the direction of our target source. 
The major observational results derived in this paper are given below:\\
$\bullet$ An inverted Y-like structure (major axis $\sim$19 pc) is investigated in the {\it Herschel} column density map, and the total of mass of this structure is determined to be $\sim$6280 M$_{\odot}$. At least two elongated filaments (having lengths $>$10 pc) are identified in the direction of the inverted Y-like structure.\\
$\bullet$ One elongated filament overlaps other one at many parts, where column densities are found to be larger than 4.5 $\times$ 10$^{21}$ cm$^{-2}$. 
In the {\it Herschel} temperature map, these filaments show a temperature range of $\sim$12.5--13.5~K.\\
$\bullet$ Based on the analysis of the $^{12}$CO and $^{13}$CO line data, molecular gas toward the inverted Y-like structure 
is studied in a velocity range of [$-$8, 0] km s$^{-1}$. 
Two elongated cloud components (around $-$1.5 and $-$4.5 km s$^{-1}$) are also found toward the inverted Y-like structure, 
and are also linked in the velocity space. \\
$\bullet$ First moment maps of $^{12}$CO and $^{13}$CO also exhibit the spatial overlapping of the two elongated filamentary cloud components along the line of sight. The morphology of the inverted Y-like structure can be explained by a combination of two different elongated filamentary clouds having a velocity separation of $\sim$3 km s$^{-1}$. These results are also supported by the distribution of the cold H\,{\sc i} gas. In other words, the inverted Y-like structure seems to contain a pair of intertwined filaments.\\ 
$\bullet$ With the knowledge of the infrared-excess sources, SF activities are found toward the inverted Y-like structure. 
The northern end of the structure contains AFGL 5142, where a cluster of YSOs is distributed within a scale of $\sim$3 pc. 
In the direction of this cluster, a very high value of surface density of YSOs (i.e., 5--240 YSOs pc$^{-2}$) is found. 
It suggests very intense ongoing SF activities (including massive star(s)) toward AFGL 5142, which is also located at one of the common zones of the two elongated clouds.
Furthermore, noticeable YSOs are also seen toward the overlapping zones of the two clouds within the inverted Y-like structure.\\  

Based on these observational outcomes, we propose a scenario of collision/interaction of two elongated filamentary clouds or flows in the site AFGL 5142, which may have triggered SF (including massive star(s)) in the site. 
\acknowledgments 
We thank the anonymous reviewer for several useful comments and suggestions. 
The research work at Physical Research Laboratory is funded by the Department of Space, Government of India. 
This work is based on data obtained as part of the UKIRT Infrared Deep Sky Survey. This publication 
made use of data products from the Two Micron All Sky Survey (a joint project of the University of Massachusetts and 
the Infrared Processing and Analysis Center / California Institute of Technology, funded by NASA and NSF), archival 
data obtained with the {\it Spitzer} Space Telescope (operated by the Jet Propulsion Laboratory, California Institute 
of Technology under a contract with NASA). 
The Canadian Galactic Plane Survey (CGPS) is a Canadian project with international partners. 
The Dominion Radio Astrophysical Observatory is operated as a national facility by the 
National Research Council of Canada. The Five College Radio Astronomy Observatory 
CO Survey of the Outer Galaxy was supported by NSF grant AST 94-20159. The CGPS is 
supported by a grant from the Natural Sciences and Engineering Research Council of Canada (NSERC). 
TB acknowledges funding from the National Natural Science Foundation of China through grant 11633005 and, support from the China Postdoctoral Foundation through grant 2018M631241.
RD acknowledges CONACyT(M\'{e}xico) for SNI grant (CVU 555629). 


\begin{thebibliography}{}
%
\bibitem[Andr{\'e} et al.(2010)]{andre10}
Andr{\'e}, P., Men'shchikov, A., Bontemps, S., et al. 2010, A\&A, 518, L102

\bibitem[Andr{\'e} et al.(2014)]{andre14}
Andr{\'e}, P., Di Francesco, J., Ward-Thompson, D., et al. 2014, in Protostars and Planets VI, ed. H. Beuther et al. (Tucson, AZ; Univ. Arizona Press), 27

\bibitem[Baker \& Burton(1979)]{baker79}
Baker, P.,~L., \& Burton, W.,~B. 1979, A\&AS, 35, 129

\bibitem[Baug et al.(2015)]{baug15}
Baug, T., Ojha, D.~K., Dewangan, L.~K., et al. 2015, MNRAS, 454, 4335

\bibitem[Benjamin et al.(2003)]{benjamin03}
Benjamin, R.~A.,Churchwell, E., Babler, B.~L., et al. 2003, PASP, 115, 953

\bibitem[Bressert et al.(2010)]{bressert10}
Bressert, E., Bastian, N., Gutermuth, R., et al. 2010, MNRAS, 409, 54

\bibitem[Burns et al.(2017)]{burns17}
Burns, R.~A., Handa, T., Imai, H., et al. 2017, MNRAS, 467, 2367

\bibitem[Brunt(2004)]{brunt04}
Brunt C., 2004, in Clemens D., Shah R., Brainerd T., eds, Proc. of ASP 
Conference 317. Milky Way Surveys: The Structure and Evolution of our Galaxy, p. 79

\bibitem[Burton et al.(1978)]{burton78}
Burton, W.,~B., Liszt, H.,~S., \& Baker, P.,~L. 1978, ApJ, 219, 67

\bibitem[Burton \& Liszt(1981)]{burton81}
Burton, W.,~B., \& Liszt, H.,~S. 1981, in Origin of Cosmic Rays, eds. G. Setti, G. Spada, \& A. W. Wolfendale, IAU Symp., 94, 227

\bibitem[Colzi et al.(2018)]{colzi18}
Colzi, L., Fontani, F., Caselli, P., Ceccarelli, C., Hily-Blant, P., \& Bizzocchi, L. 2018, A\&A, 609, 129 

\bibitem[Condon et al.(1998)]{condon98}
Condon, J.~J., Cotton, W.~D., Greisen, E.~W., et al. 1998, AJ, 115, 1693

\bibitem[Contreras et al.(2016)]{contreras16}
Contreras, Y., Garay, G., Rathborne, J.~M., \& Sanhueza,P. 2016, MNRAS, 456, 2041

\bibitem[Dewangan et al.(2015)]{dewangan15}
Dewangan, L.~K., Luna, A., Ojha, D.~K., et al.  2015, ApJ, 811, 79

\bibitem[Dewangan et al.(2016a)]{dewangan16}
Dewangan, L.~K., Ojha, D.~K., Luna, A., et al.  2016a, ApJ, 819, 66

\bibitem[Dewangan et al.(2016b)]{dewangan16b}
Dewangan, L.~K., Ojha, D.~K., Zinchenko, I., et al.  2016b, ApJ, 833, 246

\bibitem[Dewangan et al.(2017a)]{dewangan17a}
Dewangan, L.~K., Ojha, D.~K., Zinchenko, I., Janardhan, P., \& Luna, A.  2017a, ApJ, 834, 22

\bibitem[Dewangan et al.(2017b)]{dewangan17b}
Dewangan, L.~K., Ojha, D.~K., \& Baug, T. 2017b, ApJ, 844, 15

\bibitem[Dewangan et al.(2017c)]{dewangan17c}
Dewangan, L.~K., Baug, T., Ojha, D.~K., Janardhan, P., Devaraj, R., \& Luna, A. 2017c, ApJ, 845, 34

\bibitem[Dewangan et al.(2017d)]{dewangan17d}
Dewangan, L.~K., Devaraj, R., Baug, T., \& Ojha, D.~K.  2017d, ApJ, 848, 51

\bibitem[Dewangan et al.(2017e)]{dewangan17e}
Dewangan, L.~K., Ojha, D.~K., \& Zinchenko, I. 2017e, ApJ, 851, 140

\bibitem[Dewangan et al.(2018)]{dewangan18}
Dewangan, L.~K., Ojha, D.~K., Zinchenko, I., \& Baug, T. 2018, ApJ, 861, 19

\bibitem[Duarte-Cabral et al.(2011)]{duarte11}
Duarte-Cabral, A., Dobbs, C.~L., Peretto, N., Fuller, G.~A. 2011, A\&A, 528, 50


\bibitem[Evans et al.(2009)]{evans09}
Evans, N.~J., II, Dunham, M.~M., J\o{}rgensen, J.~K., et al. 2009, ApJS, 181, 321

\bibitem[Flaherty et al.(2007)]{flaherty07}
Flaherty, K.~M., Pipher, J.~L., Megeath, S.~T., et al. 2007, ApJ, 663, 1069

\bibitem[Fukui et al.(2014)]{fukui14} 
Fukui, Y., Ohama, A., Hanaoka, N., et al. 2014, ApJ, 780, 36

\bibitem[Gutermuth et al.(2009)]{gutermuth09}
Gutermuth R.~A., Megeath S.~T., Myers P.~C., et al., 2009, ApJS, 184, 18

\bibitem[Galv{\'a}n-Madrid et al.(2010)]{galvan10}
Galv{\'a}n-Madrid, R., Zhang, Q., Keto, E., et al. 2010, ApJ, 725, 17


\bibitem[Goddi \& Moscadelli(2006)]{goddi06} 
Goddi, C. \& Moscadelli, L. 2006, A\&A, 447, 577

\bibitem[Goddi et al.(2011)]{goddi11} 
Goddi, C., Moscadelli, L., \& Sanna, A. 2011, A\&A, 535, L8

\bibitem[Habe \& Ohta(1992)]{habe92} 
Habe, A., \& Ohta, K. 1992, PASJ, 44, 203 


\bibitem[Henshaw et al.(2013)]{henshaw13} 
Henshaw, J.~D., Caselli, P., Fontani, F., Jim\'{e}nez-Serra, I., Tan, J.~C., \& Hernandez, A.~K. 2013, MNRAS, 428, 3425

\bibitem[Heyer et al.(1996)]{heyer96}
Heyer, M.~H., Carpenter, J.~M., \& Ladd, E.~F. 1996, ApJ, 463, 630

\bibitem[Heyer et al.(1998)]{heyer98}
Heyer, M., Brunt, C., Snell, R., et al. 1998, ApJS, 115, 241

\bibitem[Hunter et al.(1995)]{hunter95} 
Hunter, T.~R., Testi, L., Taylor, G.~B., Tofani, G., Felli, M., \& Phillips, T.~G. 1995, A\&A, 302, 249

\bibitem[Hunter et al.(1999)]{hunter99} 
Hunter, T.~R., Testi, L., Zhang, Q., \& Sridharan, T.~K. 1999, AJ, 118, 477

\bibitem[Inoue \& Fukui(2013)]{inoue13} 
Inoue, T., \& Fukui, Y. 2013, ApJL, 774, 31

\bibitem[Kerton(2005)]{kerton05}
Kerton, C.~R. 2005, ApJ, 623, 235

\bibitem[Marsh et al.(2015)]{marsh15} 
Marsh, K.~A., Whitworth, A.~P., \& Lomax, O. 2015, MNRAS, 454, 4282

\bibitem[Marsh et al.(2017)]{marsh17} 
Marsh, K.~A., Whitworth, A.~P., Lomax, O., et al. 2017, MNRAS, 471, 2730

\bibitem[McKee \& Ostriker(2007)]{mckee07}
McKee, C.~F., \& Ostriker, E.~C. 2007, ARA\&A, 45, 565

\bibitem[Molinari et al.(2010a)]{molinari10}
Molinari, S., Swinyard, B., Bally, J., et al., 2010a, A\&A, 518, L100

\bibitem[Molinari et al.(2010b)]{molinari10b}
Molinari, S., Swinyard, B., Bally, J., et al., 2010b, PASP, 122, 314

\bibitem[Myers (2009)]{myers09} 
Myers, P.~C. 2009, ApJ, 700, 1609

\bibitem[Lawrence et al.(2007)]{lawrence07}
Lawrence, A., Warren, S.~J., Almaini, O., et al. 2007, MNRAS, 379, 1599

\bibitem[Liszt et al.(1981)]{liszt81}
Liszt, H.,~S., Burton, W.,~B., \& Bania, T.,~M. 1981, ApJ, 246, 74

\bibitem[Liu et al.(2016)]{liu16} 
Liu, T., Zhang, Q., Kim, K.-T., et al. 2016, ApJ, 824, 31

\bibitem[Nakamura et al.(2012)]{nakamura12}
Nakamura, F., Miura, T., Kitamura, Y., et al. 2012, ApJ, 746, 25

\bibitem[Nakamura et al.(2014)]{nakamura14}
Nakamura, F., Sugitani, K., Tanaka, T., et al. 2014, ApJL, 791, L23

\bibitem[Palau et al.(2011)]{palau11} 
Palau, A., Fuente, A., Girart, J.~M. et al. 2011, ApJL, 743, L32

\bibitem[Palau et al.(2013)]{palau13} 
Palau, A., Fuente, A., Girart, J.~M. et al. 2013, ApJ, 762, 120


\bibitem[Schneider et al.(2012)]{schneider12}
Schneider, N., Csengeri, T., Hennemann, M., et al. 2012, A\&A, 540, L11

\bibitem[Skrutskie et al.(2006)]{skrutskie06}
Skrutskie, M.~F., Cutri, R.~M., Stiening, R., et al. 2006, AJ, 131, 1163

\bibitem[Taylor et al.(2003)]{taylor03}
Taylor, A.~R., Gibson, S.~J., Peracaula, M., et al. 2003, AJ, 125, 3145

\bibitem[Qiu et al.(2008)]{qiu08}
Qiu, K., Zhang, Q., Megeath, S.~T., et al. 2008, ApJ, 685, 1005

\bibitem[Whitney et al.(2011)]{whitney11}
Whitney, B., Benjamin, R., Meade, M., et al. 2011, BAAS, 43, 241.16

\bibitem[Williams et al.(2018)]{williams18} 
Williams, G.~M., Peretto, N., Avison, A., Duarte-Cabral, A., \& Fuller, G.~A. 2018, A\&A, 613, 11

\bibitem[Wright et al.(2010)]{wright10}
Wright, E.~L., Eisenhardt, P.~R.~M., Mainzer, A.~K., et al. 2010, AJ, 140, 1868

\bibitem[Zhang et al.(2007)]{zhang07} 
Zhang, Q., Hunter, T.~R., Beuther, H. et al. 2007, ApJ, 658, 1152

\end{thebibliography}
 \end{document}